\documentclass[lettersize,journal]{IEEEtran}
\pdfoutput=1
\usepackage{amsmath,amsfonts,amsthm,amssymb}
\usepackage{algorithmic}
\usepackage{algorithm}
\usepackage{array}
\usepackage[caption=false,font=normalsize,labelfont=sf,textfont=sf]{subfig}
\usepackage{textcomp}
\usepackage{stfloats}
\usepackage{url}
\usepackage{verbatim}
\usepackage{graphicx}
\usepackage{cite}
\usepackage{bm}
\usepackage{xcolor}
\usepackage{orcidlink}
\hypersetup{hidelinks}
\usepackage{pstricks}
\begin{document}

\title{Hybrid Iterative Detection for OTFS: Interplay between Local L-MMSE and Global Message Passing}

\author{Ruohai~Yang$^{\orcidlink{0009-0006-1355-9366}}$,
        Shuangyang~Li$^{\orcidlink{0000-0002-9503-4865}}$,~\IEEEmembership{Member,~IEEE},
        Han~Yu$^{\orcidlink{0000-0002-8489-2266}}$,~\IEEEmembership{Member,~IEEE},
        \\
        Zhiqiang~Wei$^{\orcidlink{0000-0003-3400-5590}}$,~\IEEEmembership{Member,~IEEE},
        Kai~Wan$^{\orcidlink{0000-0003-4671-3287}}$,~\IEEEmembership{Member,~IEEE},
        and Giuseppe~Caire$^{\orcidlink{0000-0002-7749-1333}}$,~\IEEEmembership{Fellow,~IEEE}
\thanks{R.~Yang and K.~Wan are with the School of Electronic Information and Communications, Huazhong University of Science and Technology, Wuhan 430074, China (e-mails: \{yngrh,kai\_wan\}@hust.edu.cn). 

S.~Li, H.~Yu, and Giuseppe~Caire are with the Faculty of Electrical Engineering and Computer Science, Technical University of Berlin, 10587 Berlin, Germany (e-mail: shuangyang.li@tu-berlin.de, han.yu.1@campus.tu-berlin.de, caire@tu-berlin.de). 

Z.~Wei is with the School of Mathematics and Statistics, Xi'an Jiaotong University, Xi'an 710049, China (e-mail: zhiqiang.wei@xjtu.edu.cn).}
}

\maketitle

\begin{abstract}
Orthogonal time frequency space (OTFS) modulation has emerged as a robust solution for high-mobility wireless communications. However, conventional detection algorithms, such as linear equalizers and message passing (MP) methods, either suffer from noise enhancement or fail under complex doubly-selective channels, especially in the presence of fractional delay and Doppler shifts. In this paper, we propose a hybrid low-complexity iterative detection framework that combines linear minimum mean square error (L-MMSE) estimation with MP-based probabilistic inference. 
The key idea is to apply a new delay-Doppler (DD) commutation precoder (DDCP) to the DD domain signal vector, such that the resulting effective channel matrix exhibits a structured form with several locally dense blocks that are sparsely inter-connected. This precoding structure enables a hybrid iterative detection strategy, where a low-dimensional L-MMSE estimation is applied to the dense blocks, while MP is utilized to exploit the sparse inter-block connections. Furthermore, we provide a detailed complexity analysis, which shows that the proposed scheme incurs lower computational cost compared to the full-size L-MMSE detection. The simulation results of convergence performance confirm that the proposed hybrid MP detection achieves fast and reliable convergence with controlled complexity. In terms of error performance, simulation results demonstrate that our scheme achieves significantly better bit error rate (BER) under various channel conditions. Particularly in multipath scenarios, the BER performance of the proposed method closely approaches the matched filter bound (MFB), indicating its near-optimal error performance. 
\end{abstract}


\section{Introduction}
The explosive growth of high-mobility wireless applications~\cite{Zhou2020}, such as high-speed railways, vehicle-to-everything (V2X) communications, and unmanned aerial vehicles (UAVs), places unprecedented demands on wireless communication systems in terms of reliability, latency, and spectral efficiency \cite{Matthaiou2021}. However, traditional multicarrier schemes, particularly orthogonal frequency division multiplexing (OFDM), become increasingly ineffective in such dynamic scenarios. The core limitation of OFDM stems from its assumption of quasi-static channels over one frame duration, which breaks down under large Doppler spreads, resulting in potentially severe inter-carrier interference (ICI) and severe performance degradation \cite{Hwang2009,Tiejun2006}.

Orthogonal Time Frequency Space (OTFS) modulation has recently emerged as a promising candidate for robust communications over doubly selective channels \cite{Hadani2017,Wei2021}. Different from OFDM, which transmits symbols in the time-frequency (TF) domain, OTFS maps information symbols onto the delay-Doppler (DD) domain and converts these symbols to the TF domain via the inverse symplectic finite Fourier transform (ISFFT). This transformation enables OTFS to exploit the inherent sparsity and quasi-static nature of DD domain wireless channels~\cite{Gaudio2022}. Consequently, OTFS can potentially capture the full diversity of the time-varying channel within each frame and achieve superior robustness against both delay and Doppler spreads \cite{Li2021,Raviteja2020}. Furthermore, the structured and often sparse nature of DD domain channels facilitates the design of low-overhead channel estimation \cite{Raviteja2019,Yuan2021}, as well as advanced detection methods \cite{Raviteja2018,Surabhi2020,Pandey2021,Wang2024,li2022crossdomain}.

Despite these advancements, achieving robust and low-complexity detection for OTFS under general channel conditions remains a significant challenge. Numerous works have been proposed to exploit the inherent sparsity and structure of the DD domain. Among them, message passing (MP) \cite{Raviteja2018} detection has received considerable attention due to its capability to efficiently handle sparse interference structures via probabilistic graphical models. By representing the OTFS input-output relation as a sparse factor graph, the MP algorithm approximates interference using Gaussian messages and iteratively refines symbol estimates through forward and backward message updates. 
However, the performance of MP may degrade due to dense interference from fractional delay and Doppler shifts, or due to short cycles in the factor graph. 
Another representative strategy is the cross-domain iterative detection (CDID) framework~\cite{li2022crossdomain}, which exploits both the sparsity of the DD domain and the approximately diagonalization of the channel in the time domain. CDID alternates between linear estimation in the time domain and non-linear detection in the DD domain, exchanging soft information across domains via inverse SFFT and SFFT. However, the CDID method introduces additional computational overhead due to multiple domain conversions. In parallel, low complexity linear equalizations, including zero-forcing (ZF) and linear minimum mean square error (L-MMSE) equalizers, have been explored to reduce algorithmic complexity~\cite{Surabhi2020}. Although these detectors require low complexity, their detection performances are usually limited. 

The main contribution of this paper is to propose a new hybrid iterative detection framework that non-trivially combines the L-MMSE and MP algorithms. The core idea is to decouple the detection process into two interacting parts: one focuses on interference suppression through low-complexity linear estimation, while the other refines symbol-wise posterior distributions via iterative MP. More precisely, the technical contribution contains the following points. 

 \begin{itemize}
    \item 
    To support efficient and robust detection under practical OTFS channel conditions, we apply a structured delay-Doppler commutation precoder (DDCP) to the DD domain signal. This precoding transforms the effective channel matrix into a block-sparse pattern, where each non-empty block is dense and the sparsity lies across blocks. This design enables separate treatment of local and global interference, laying the foundation for modular hybrid detection.

    \item 
    Building on the block-sparse structure, we develop a hybrid and tightly coupled detector, in which L-MMSE acts locally and MP acts globally. The core idea is to decouple the detection into two interacting parts: (i) a local L-MMSE module that operates on each dense block to suppress strong intra-block interference with low complexity; and (ii) a global MP module that runs on the inter-block factor graph to reconcile residual cross-block interference. This design is particularly effective under fractional Doppler conditions, where conventional MP fails due to the loss of sparsity. To ensure stable convergence and avoid excessive iterations, we incorporate a damping mechanism to control update dynamics and introduce a stopping criterion based on \textit{a posteriori} certainty, promising fast and reliable convergence with controlled complexity.

    \item 
    We conduct comprehensive theoretical analysis and numerical evaluations to validate the performance of the proposed detection algorithm. Specifically, complexity analysis, convergence studies, and extensive simulations demonstrate that our scheme achieves near-optimal bit-error rate (BER) performance while maintaining a manageable computational cost. Moreover, it shows that the proposed design is robust under dense channel conditions induced by fractional delay shifts and is compatible with both ISFFT/SFFT-based and inverse Zak transform (IZT)/ZT-based OTFS system implementations.
\end{itemize}

\textbf{Notations:} The blackboard bold letter $\mathbb{C}$ denotes the complex number field; bold-face capitals and lower-case letters are used to define   matrices and  vectors, respectively; 
 $(*)^\mathrm{T}$, $(*)^\star$, $(*)^{-1}$ and $(*)^\mathrm{H}$ denote the transpose, the conjugate, the inverse and the Hermitian operations for a matrix, respectively; $\otimes$ denotes the Kronecker product operator; $\mathrm{vec}(*)$ denotes the vectorization operation; $\mathrm{Pr}(*)$ denotes the probability of an event; $\propto $ represents that both sides of the equation are multiplicatively connected to a constant; the big-O notation $\mathcal{O}(\cdot)$ asymptotically describes the order of computational complexity; the superscripts \(\mathrm{a} \), \( \mathrm{p} \), and \(\mathrm{e} \) are used to indicate \textit{a priori}, \textit{a posteriori}, and extrinsic information, respectively.

\section{System Model}
\subsection{General DD Domain Representation}
\begin{figure}
\centering
\includegraphics[width=3in]{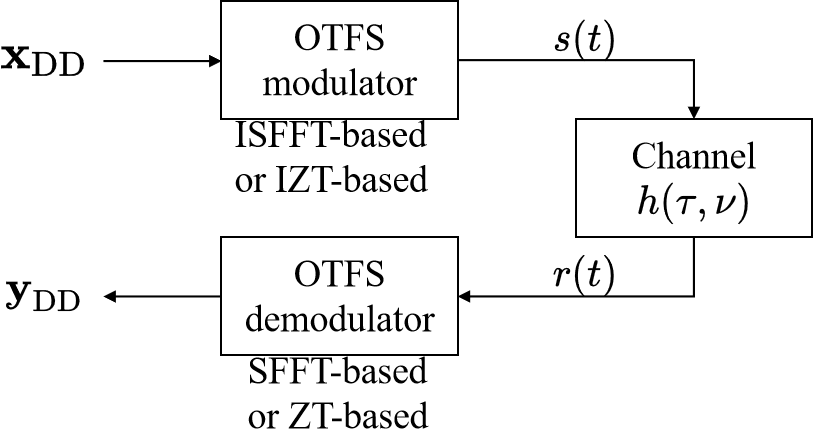}
\caption{The general OTFS system model of consideration.}
\label{OTFS_sys}
\end{figure}

We consider the OTFS transceiver architecture illustrated in Fig.~\ref{OTFS_sys}. Several practical implementations of the OTFS modulator and demodulator have been proposed, among which the ISFFT/SFFT and IZT/ZT frameworks are the most widely used~\cite{RezazadehReyhani2018,Lampel2022}. 
For clarity, we use the superscripts \( \mathrm{ISFFT} \) and \( \mathrm{IZT} \) to denote the corresponding effective time-domain and DD-domain channels.
This section focuses on a general model of the OTFS system. The specific implementation forms of the modulator and demodulator, such as those based on ISFFT/SFFT or IZT/ZT, are not considered here and will be discussed in Sections II-B and II-C, respectively.

Let \( N \) and \( M \) denote the number of Doppler bins/time slots and delay bins/subcarriers, respectively. The transmitted information symbols are drawn from a finite constellation set \( \mathbb{A} = \{a_1, a_2, \cdots, a_Q\} \). 

At the transmitter side, the DD domain symbol vector \( \mathbf{x}_\text{DD} \in \mathbb{A}^{MN} \) is directly formed by selecting symbols from the constellation set \( \mathbb{A} \). $\mathbf{x}_\text{DD}$ is then converted into a continuous-time baseband signal \( s(t) \) by the OTFS modulator. This modulation process effectively spreads the DD symbols across both time and frequency, offering resilience against doubly selective fading. 

An OTFS signal is transmitted over a time-varying channel, whose response can be fully characterized by its DD domain representation as given by \cite{Wireless2011}, 
\begin{equation}
h\left(\tau,\nu\right)=\sum_{i=1}^P h_i \delta\left(\tau-\tau_i\right)\delta\left(\nu-\nu_i\right).
\label{eq1}
\end{equation}
Here, \( P \) denotes the number of resolvable paths, while \( h_i \), \( \tau_i \), and \( \nu_i \) represent the path gain, delay shift, and Doppler shift associated with the \( i \)-th path, respectively, where
\begin{equation}
\tau_i=\frac{l_i +\iota_i}{M\Delta f},\quad\mathrm{and}\quad\nu_i=\frac{k_i+\kappa_i}{NT}.
\label{eq2}
\end{equation}
Note that \(\Delta f\) denotes the spacing between adjacent subcarriers. The integers \(l_i\) and \(k_i\) are the integer parts of the delay and Doppler indices, respectively, with \(l_i \in \{0,1,\ldots,M-1\}\) and \(k_i \in \{-\tfrac{N}{2},\ldots,\tfrac{N}{2}\}\). The quantity \(\kappa_i \in [-\tfrac{1}{2},\,\tfrac{1}{2})\) is the fractional Doppler offset from the nearest Doppler grid point, and \(\iota_i \in [0,1)\) is the fractional delay within one delay bin. Under the wideband assumption, the typical sampling interval \( 1/ (M\Delta f) \) in the time domain is sufficiently small; thus the fractional delay component for each path has been neglected in some OTFS works such as \cite{Raviteja20182}, i.e., \( \iota_i = 0 \) for \( 1 \le i \le P \). However, in this paper, both integer and fractional delay cases are considered for a comprehensive understanding.

At the receiver side, the received signal can then be expressed as
\begin{equation}
r\left(t\right)=\int\int h\left(\tau,\nu\right)s\left(t-\tau\right)e^{j2\pi\nu\left(t-\tau\right)}\mathrm d\tau \mathrm d\nu + n\left(t\right),
\label{eq3}
\end{equation}
where \( n(t) \) denotes the additive white Gaussian noise (AWGN)   with a one-sided power spectral density (PSD) of \( N_0 \). Similar to the transmitter side, the received symbol vector in the  DD domain, \( \mathbf{y}_\mathrm{DD} \in \mathbb{C}^{MN} \), is obtained via an OTFS demodulator, which typically involves the SFFT and the ZT. 

We note that the OTFS modulator and demodulator are linear with respect to the DD domain symbols, and the physical channel is linear time-varying. Hence, regardless of the specific realization, the end-to-end input–output mapping in the DD domain remains linear. We therefore model it by an equivalent DD domain channel matrix \(\mathbf{H}_{\mathrm{DD}}\in\mathbb{C}^{MN\times MN}\) acting on the transmit vector \(\mathbf{x}_{\mathrm{DD}}\), with an additive noise vector \(\mathbf{n}_{\mathrm{DD}}\). Then we have 
\begin{equation}
\mathbf{y}_\mathrm{DD} = \mathbf{H}_\mathrm{DD} \mathbf{x}_\mathrm{DD} + \mathbf{n}_\mathrm{DD},
\label{eq4}
\end{equation}
where the one-sided PSD of \( \mathbf{n}_\mathrm{DD} \) is \( N_0 \). Different implementations of OTFS result in different forms of the equivalent DD domain channel matrix \( \mathbf{H}_\mathrm{DD} \). Specifically, the ISFFT/SFFT-based channel \( \mathbf{H}_\mathrm{DD}^\text{ISFFT} \) will be introduced in Section~II-B, while the IZT/ZT-based channel \( \mathbf{H}_\mathrm{DD}^\text{IZT} \) will be discussed in Section~II-C.

\subsection{ISFFT/SFFT-Based Channel \( \mathbf{H}_\mathrm{DD}^\mathrm{ISFFT} \)}
First, we describe the OTFS modulator-demodulator based on the ISFFT/SFFT. This implementation yields a concise input-output relation when operating over channels with integer delay shifts (i.e., \( \iota_i = 0 \) for \(1 \le i \le P\)), while allowing either integer or fractional Doppler shifts.

At the transmitter side, the data stream \( \mathbf{x}_\mathrm{DD} \) is modulated by the OTFS transmitter discussed in \cite{Hadani2017}, which involves the ISFFT followed by the Heisenberg transform. Assuming a rectangular pulse in the Heisenberg transform, the resultant time-domain baseband OTFS signal in vector form, \( \mathbf{s} \in \mathbb{C}^{MN} \), is given by \cite{Raviteja20192}, 
\begin{equation}
\mathbf{s} = \left( \mathbf{F}_N^\mathrm{H} \otimes \mathbf{I}_M \right) \mathbf{x}_\text{DD},
\label{eq5}
\end{equation}
where \( \mathbf{F}_N \) is the discrete Fourier transform (DFT) matrix of size \( N \times N \), and \( \mathbf{I}_M \) is the identity matrix of size \( M \times M \).

With a reduced cyclic prefix (RCP) structure, the time-domain received signal vector after removing the CP is 
\begin{equation}
\mathbf{r} = \mathbf{H}^\mathrm{ISFFT}_\mathrm{T} \mathbf{s} + \mathbf{n}_\mathrm{T},
\label{eq6}
\end{equation}
where \( \mathbf{n}_\mathrm{T} \) is the corresponding time domain AWGN. In (\ref{eq6}), \( \mathbf{H}^\mathrm{ISFFT}_\mathrm{T} \) denotes the effective channel matrix in the time domain, of size \( MN \times MN \), and is given by \cite{Raviteja20192},
\begin{equation}
\mathbf{H}^\mathrm{ISFFT}_\mathrm{T} = \sum_{i=1}^P h_i e^{\frac{-j2\pi(k_i+\kappa_i)l_i}{MN}} \bm{\Delta}^{k_i+\kappa_i} \bm{\Pi}_{MN}^{l_i},
\label{eq7}
\end{equation}
where \( \bm{\Pi}_{MN} \) is the permutation matrix (forward cyclic shift) of size \( MN \times MN \), defined as
\begin{equation}
\bm{\Pi}_{MN} = \begin{bmatrix}0 & \cdots & 0 & 1\\1 & \ddots & 0 & 0\\\vdots & \ddots & \ddots & \vdots\\0 & \cdots & 1 & 0\end{bmatrix}_{MN \times MN},
\label{eq8}
\end{equation}
and \( \bm{\Delta} = \mathrm{diag}\{\alpha^0, \alpha^1, \ldots, \alpha^{MN-1}\} \) is a diagonal matrix with \( \alpha \triangleq e^{\frac{j2\pi}{MN}} \). 

Then, by applying OTFS demodulation with respect to the rectangular pulse, which includes the SFFT and the Wigner transform, the received symbol vector is given by
\begin{equation}
\mathbf{y}_\mathrm{DD} = \left( \mathbf{F}_N \otimes \mathbf{I}_M \right) \mathbf{r}.
\label{eq9}
\end{equation}
By combining equations (\ref{eq4}), (\ref{eq5}), (\ref{eq6}), and (\ref{eq9}), the equivalent channel matrix in the DD domain, \( \mathbf{H}^\mathrm{ISFFT}_\mathrm{DD} \), and the corresponding AWGN noise vector \( \mathbf{n}_\mathrm{DD} \) can be expressed as
\begin{equation}
\mathbf{H}^\mathrm{ISFFT}_{\mathrm{DD}}=\left(\mathbf{F}_N\otimes\mathbf{I}_M\right)\mathbf{H}^\mathrm{ISFFT}_\mathrm{T}\left(\mathbf{F}_N^\mathrm{H}\otimes\mathbf{I}_M\right),
\label{eq10}
\end{equation}
and 
\begin{equation}
\mathbf{n}_{\mathrm{DD}}=\left(\mathbf{F}_N\otimes\mathbf{I}_M\right)\mathbf{n}_\mathrm{T}.
\label{eq11}
\end{equation}

\subsection{IZT/ZT-Based Channel \( \mathbf{H}_\mathrm{DD}^\mathrm{IZT} \)}
Second, we discuss the OTFS modulator-demodulator structure based on the IZT and the ZT, as described in~\cite{Chong2024}. This implementation is advantageous in deriving a concise input-output relation in the presence of fractional delay indices.

Consider a ZT-based OTFS implementation employing rectangular windows \cite{li20232}. After passing \( \mathbf{x}_\mathrm{DD} \) through the IZT module, the time-domain OTFS symbol vector \( \mathbf{s} \) is still given by (\ref{eq5}). Inserting the RCP to \( \mathbf{s} \), the time-domain transmitted symbol vector \( \tilde{\mathbf{s}} \in \mathbb{C}^{MN + L_\mathrm{CP}} \) is 
\begin{equation}
\tilde{\mathbf{s}} = \mathbf{A}_\mathrm{CP} \mathbf{s} = \mathbf{A}_\mathrm{CP} \left( \mathbf{F}_N^\mathrm{H} \otimes \mathbf{I}_M \right) \mathbf{x}_\text{DD},
\label{eq12}
\end{equation}
where \( \mathbf{A}_\mathrm{CP} \) is the RCP insertion matrix. Specifically, \( \mathbf{A}_\mathrm{CP} \triangleq \left[ \mathbf{G}_\mathrm{CP}, \mathbf{I}_{MN} \right]^\mathrm{T} \), where \( \mathbf{G}_\mathrm{CP} \in \mathbb{C}^{MN \times L_\mathrm{CP}} \) consists of the last \( L_\mathrm{CP} \) columns of the identity matrix \( \mathbf{I}_{MN} \) \cite{RezazadehReyhani2018}.

Let $p\left(t\right)$ be the transmit pulse shaping function. In this paper, we choose the sinc function as the pulse shaping function, 
\begin{equation}
p(t)=\frac{1}{\sqrt{T_s}}\text{sinc}(\frac{t}{T_s}),
\label{eq_a2}
\end{equation}
where \( T_s = \frac{T}{M} \) is the delay resolution and $\text{sinc}(x)=\sin{(\pi x)}/(\pi x)$. The continuous-time transmitted signal after pulse shaping is 
\begin{equation}
s(t) = \sum_{n=-L_\mathrm{CP}+1}^{MN} \tilde{s}[n] p(t - nT_s),
\label{eq13}
\end{equation}
and \( \tilde{s}[n] \) denotes the \( n \)-th entry of \( \tilde{\mathbf{s}} \). By~\eqref{eq3}, the received signal \( r(t) \) is 
\begin{equation}
\begin{aligned}
r(t) &= \sum_{i=1}^{P} h_i e^{j2\pi \nu_i (t - \tau_i)} s(t - \tau_i) + n(t) \\
     &= \sum_{i=1}^{P} \sum_{n=-L_\mathrm{CP}+1}^{MN} \negmedspace\negmedspace\negmedspace h_i e^{j2\pi \nu_i (t - \tau_i)} \tilde{s}[n] p(t - nT_s - \tau_i) + n(t).
\end{aligned}
\label{eq14}
\end{equation}

Passing the received signal through a matched filter with the receive pulse \( p^\star(t) \) yields a set of sufficient statistics \( \tilde{\mathbf{r}} \) for detection. For convenience, we define
\begin{equation}
G_i[m,n] \triangleq h_i e^{j 2 \pi n \nu_i T_{s}} A_{p}^{*}\left( (n - m) T_{s} + \tau_i, \nu_i \right),
\label{eq15}
\end{equation}
as the effective time-domain channel coefficient describing the contribution of the \( n \)-th transmit symbol \( \tilde{s}[n] \) to the \( m \)-th received symbol \( \tilde{r}[m] \) via the \( i \)-th propagation path. Here, \( A_p(\tau, \nu) \) denotes the ambiguity function of the shaping pulse \( p(t) \), defined as
\begin{equation}
A_p(\tau, \nu) \triangleq \int_{-\infty}^{\infty} p(t) p^\star(t - \tau) e^{-j 2 \pi \nu (t - \tau)} \mathrm{d} t.
\label{eq16}
\end{equation}
The received vector \( \tilde{\mathbf{r}} \) can then be expressed in vector form as
\begin{equation}
\tilde{\mathbf{r}} = \sum_{i=1}^{P} \mathbf{G}_i \tilde{\mathbf{s}} + \tilde{\mathbf{n}}_\mathrm{T},
\label{eq17}
\end{equation}
where \( \tilde{\mathbf{n}}_\mathrm{T} \) is the discrete-time AWGN vector, and \( \mathbf{G}_i \) denotes the effective time-domain channel matrix corresponding to the \( i \)-th path, whose \( (m,n) \)-th element is \( G_i[m,n] \).

Let \( \mathbf{R}_\mathrm{CP} \) denote the RCP removal matrix of size \( MN \times (MN + L_\mathrm{CP}) \), obtained by removing the first \( L_\mathrm{CP} \) rows of \( \mathbf{I}_{MN + L_\mathrm{CP}} \). Accordingly, the time-domain OTFS symbol vector \( \mathbf{r} \) after removing the RCP is given by
\begin{equation}
\mathbf{r} = \mathbf{R}_\mathrm{CP} \left( \sum_{i=1}^{P} \mathbf{G}_i \mathbf{A}_\mathrm{CP} \mathbf{s} \right) + \mathbf{n}_\mathrm{T}.
\label{eq18}
\end{equation}

Finally, by transforming the time-domain vector \( \mathbf{r} \) into the DD domain using the ZT, the corresponding DD domain input-output relation can be expressed as
\begin{equation}
\mathbf{y}_\mathrm{DD} = \sum_{i=1}^{P} (\mathbf{F}_N \otimes \mathbf{I}_M) \mathbf{R}_\mathrm{CP} \mathbf{G}_i \mathbf{A}_\mathrm{CP} (\mathbf{F}_N^\mathrm{H} \otimes \mathbf{I}_M) \mathbf{x}_\mathrm{DD} + \mathbf{n}_\mathrm{DD}.
\label{eq19}
\end{equation}

Thus, the effective DD domain channel matrix \( \mathbf{H}^\mathrm{IZT}_\mathrm{DD} \) for this implementation can be represented as
\begin{equation}
\mathbf{H}^\mathrm{IZT}_\mathrm{DD} = (\mathbf{F}_N \otimes \mathbf{I}_M) \mathbf{R}_\mathrm{CP} \left( \sum_{i=1}^{P} \mathbf{G}_i \right) \mathbf{A}_\mathrm{CP} (\mathbf{F}_N^\mathrm{H} \otimes \mathbf{I}_M).
\label{eq20}
\end{equation}


For simplicity, we use $\mathbf{H}_\mathrm{DD}$ to denote the equivalent DD domain channel matrix throughout the rest of this paper. The specific form, either $\mathbf{H}^\mathrm{ISFFT}_\mathrm{DD}$ based on (\ref{eq10}) or $\mathbf{H}^\mathrm{IZT}_\mathrm{DD}$ based on (\ref{eq20}) will only be specified in our numerical results.

\section{The proposed DDCP for OTFS System}
In this section, we introduce a precoding operation of the signal vector and channel matrix for the OTFS system, referred to as the DDCP. A similar commutation operation has been used in \cite{Ale2025commutation} to transform the channel matrix into a Kronecker-product structure, thereby facilitating analytical derivations and efficient optimization. For the considered OTFS system, we apply the proposed DDCP to effectively exchange the roles of delay and Doppler shifts in forming the equivalent DD domain channel matrix. To the best of our knowledge, such a precoding approach has not been applied in the literature of OTFS. 

For the ease of presentation, in this section, we only consider the equivalent channel matrix \( \mathbf{H}_\mathrm{DD} \) generated by the ISFFT and SFFT, i.e., $\mathbf{H}^\mathrm{ISFFT}_\mathrm{DD}$. Note that this can be directly applicable to $\mathbf{H}^\mathrm{IZT}_\mathrm{DD}$. Recall \cite{HONG202247}, the equivalent channel matrix $\mathbf{H}_\mathrm{DD}$ can be divided into $N\times N$ blocks with each block of size $M\times M$. From (\ref{eq7}) and (\ref{eq10}), we observe that \( \mathbf{H}_\mathrm{DD} \) has intra-block sparsity with integer delay shifts and inter-block sparsity with integer Doppler shifts. The ISFFT/SFFT-based implementation of the OTFS system cannot actually be applied with fractional delay shifts. We have two main motivations for proposing the   DDCP. First, in practical OTFS systems, especially in wideband scenarios, achieving high delay resolution is generally more feasible than achieving high Doppler resolution. Consequently, communication settings with integer delay shifts and fractional Doppler shifts are commonly considered. Second, the locally dense block can be diagonalized using low-dimensional matrix operations, while the inter-block sparsity can be exploited by MP globally. In particular, each dense block can be efficiently equalized via linear equalizers such as L-MMSE. Therefore, in the scenarios involving integer delays and fractional Doppler shifts, we are motivated to apply the DDCP to exchange the roles of delay and Doppler indices in constructing the equivalent DD domain channel matrix. This precoding operation transforms the channel structure from exhibiting intra-block sparsity to exhibiting inter-block sparsity.

For simplicity, we denote the original transmit and receive vectors as \( \mathbf{x}_\mathrm{DD} = \mathrm{vec}(\mathbf{X}_\mathrm{DD}) \) and \( \mathbf{y}_\mathrm{DD} = \mathrm{vec}(\mathbf{Y}_\mathrm{DD}) \), and the precoded transmit and receive vectors as \( \mathbf{x} \) and \( \mathbf{y} \), respectively. This precoding operation corresponds to a change in the vectorization direction. We can represent \( \mathbf{x} \) and \( \mathbf{y} \) as \( \mathbf{x} = \mathrm{vec}(\mathbf{X}_\mathrm{DD}^\mathrm{T}) \) and \( \mathbf{y} = \mathrm{vec}(\mathbf{Y}_\mathrm{DD}^\mathrm{T}) \). As shown in \cite{xu2018}, the commutation matrix \( \mathbf{K}_{M,N} \) can be used to express the relationship between the original and precoded vectors, i.e.,
\begin{equation}
\mathbf{x} = \mathbf{K}_{M,N} \mathbf{x}_\mathrm{DD},\quad \mathbf{y} = \mathbf{K}_{M,N} \mathbf{y}_\mathrm{DD},
\label{eq21}
\end{equation}
where
\begin{equation}
\mathbf{K}_{M,N} = \sum_{i=1}^{M} \sum_{j=1}^{N} (\mathbf{e}_j \otimes \mathbf{e}_i)(\mathbf{e}_i \otimes \mathbf{e}_j)^\mathrm{T},
\label{eq22}
\end{equation}
and \( \mathbf{e}_i ,\mathbf{e}_j \) are orthonormal basis vectors of length \( M \) and \( N \), respectively. Each \( \mathbf{e}_i \in \mathbb{C}^M \) is a vector with $1$ at the $i$-th position and 0 elsewhere, and \( \mathbf{e}_j \) is defined similarly. 

The commutation matrix \( \mathbf{K}_{M,N} \) is an orthogonal matrix, satisfying \( \mathbf{K}_{M,N}^\mathrm{T} = \mathbf{K}_{M,N}^{-1} \). Based on (\ref{eq4}) and (\ref{eq21}), the input-output relation with DDCP can be written as
\begin{equation}
\begin{aligned}
\mathbf{y} &= \mathbf{H} \mathbf{x} + \mathbf{n},
\end{aligned}
\label{eq23}
\end{equation}
where \( \mathbf{H} = \mathbf{K}_{M,N} \mathbf{H}_\mathrm{DD} \mathbf{K}_{M,N}^{-1} \) denotes the effective DD domain channel matrix with DDCP, and \( \mathbf{n} = \mathbf{K}_{M,N} \mathbf{n}_\mathrm{DD} \) is the corresponding noise vector following the same distribution as \( \mathbf{n}_\mathrm{DD} \) because \( \mathbf{K}_{M,N} \) is an orthogonal matrix.

\begin{figure*}
\centering
\subfloat[$\mathbf{H}_\text{DD}$]{\includegraphics[width=3.5in]{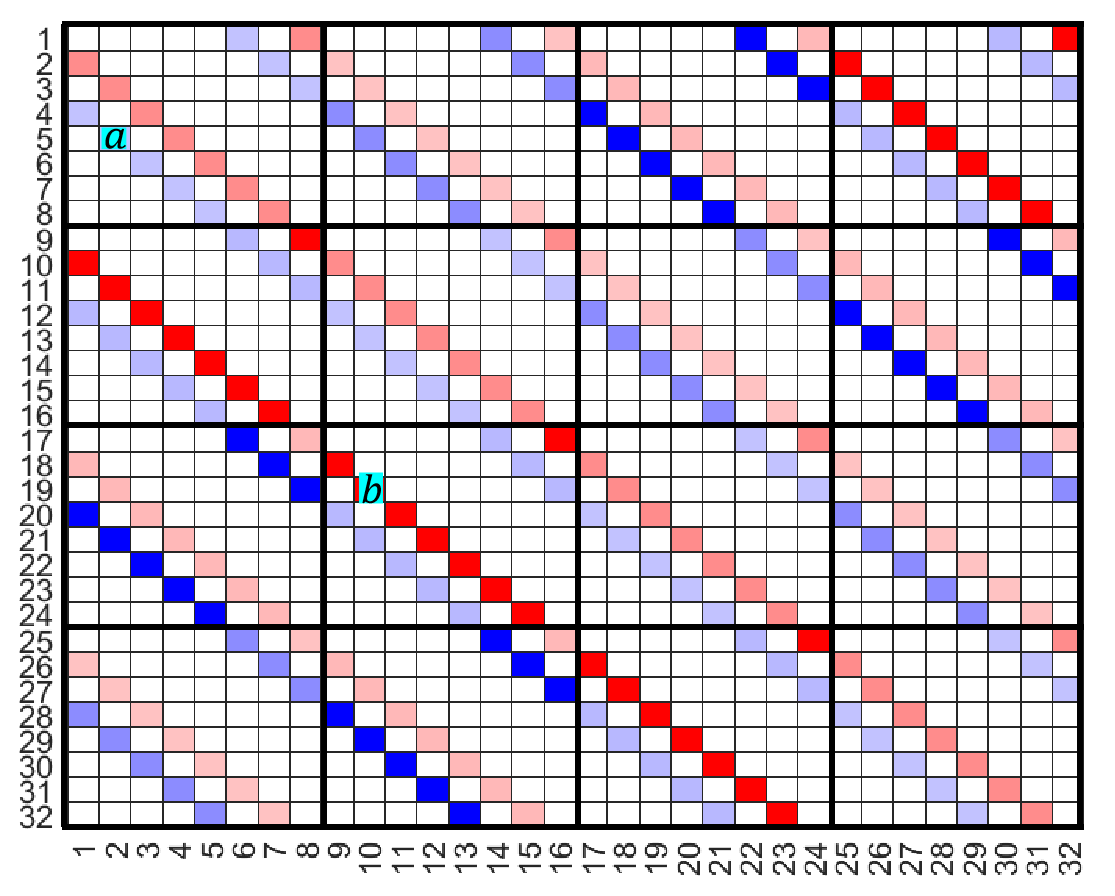}}%
\subfloat[$\mathbf{H}$]{\includegraphics[width=3.5in]{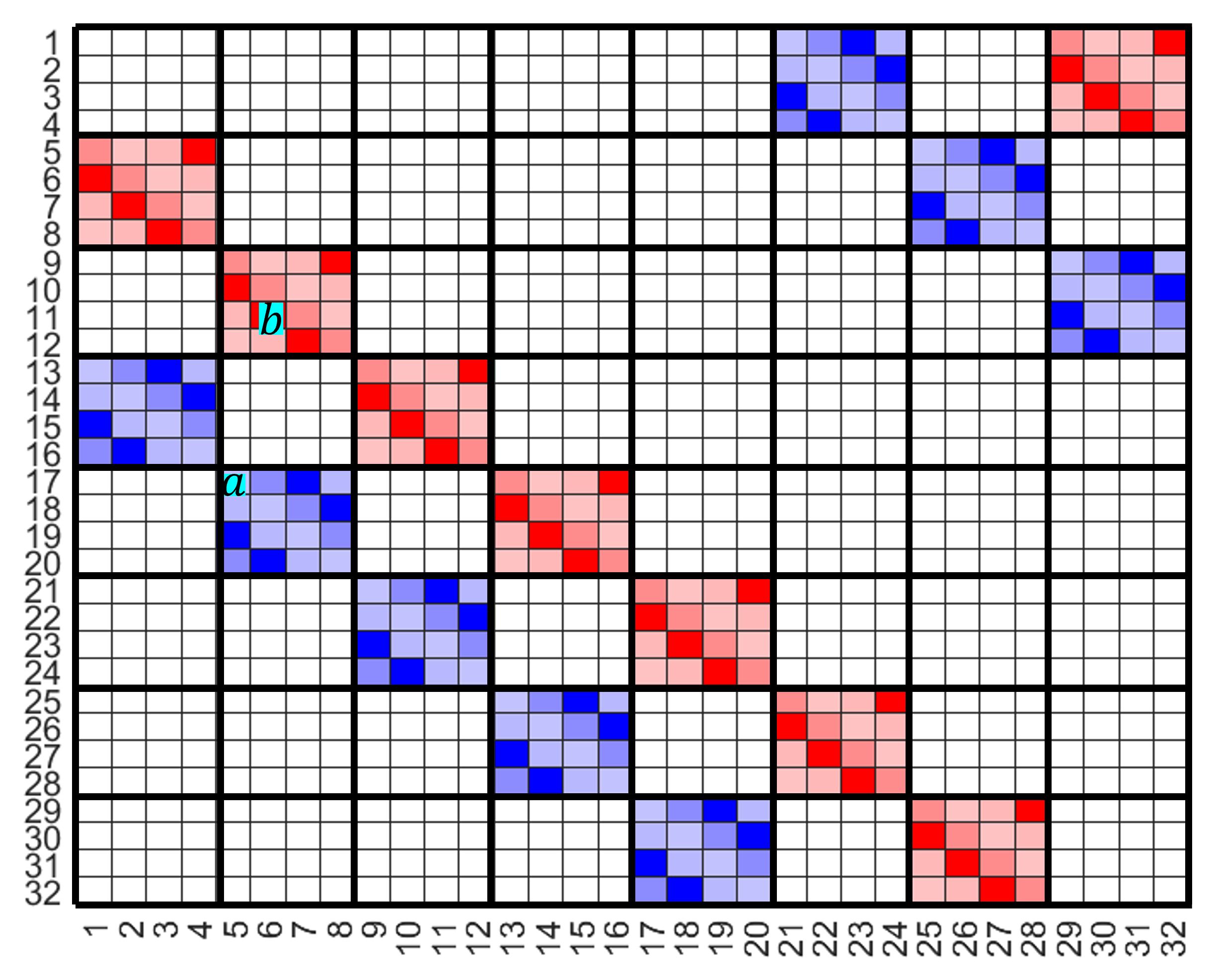}}%
\caption{Equivalent channel matrix and its counterpart with DDCP under a two-path scenario with $M=8$ and $N=4$. For the first path, $l_1=1$, $k_1=0.7$, and $h_1=1$, represented using a red color gradient. For the second path, $l_2=3$, $k_2=-1.7$, and $h_2=1$, represented using a blue color gradient. (a) has $N\times N$ blocks and shows intra-block sparsity; (b) has $M\times M$ blocks and shows inter-block sparsity. }
\label{two_H}
\end{figure*}

 An illustrative example of \( \mathbf{H}_\mathrm{DD} \) and its precoded counterpart with DDCP \( \mathbf{H} \) is given in Fig.~\ref{two_H}. $\mathbf{H}_\mathrm{DD}$ can be divided into $N\times N$ blocks with each block of size $M\times M$, while $\mathbf{H}$ can be divided into $M\times M$ blocks with each block of size $N\times N$. We use the notation \( \mathbf{H}_\mathrm{DD}^{i,j} \) to denote the \((i,j)\)-th block of \( \mathbf{H}_\mathrm{DD} \) with \(1\le i,j\le N\), and \( \mathbf{H}^{u,v} \) to denote the \((u,v)\)-th block of \( \mathbf{H} \) with \(1\le u,v\le M\).
 Thus, we can represent $\mathbf{H}_\mathrm{DD}$ and $\mathbf{H}$ as 
\begin{equation}
\mathbf{H}_\mathrm{DD}=\begin{bmatrix}
 \mathbf{H}_\mathrm{DD}^{1,1} & \cdots  & \mathbf{H}_\mathrm{DD}^{1,N}\\
\vdots   & \ddots  & \vdots \\
\mathbf{H}_\mathrm{DD}^{N,1}  & \cdots  & \mathbf{H}_\mathrm{DD}^{N,N}
\end{bmatrix},\ \mathbf{H}=\begin{bmatrix}
 \mathbf{H}^{1,1} & \cdots  & \mathbf{H}^{1,M}\\
\vdots   & \ddots  & \vdots \\
\mathbf{H}^{M,1}  & \cdots  & \mathbf{H}^{M,M}
\label{eq_a1}
\end{bmatrix}.
\end{equation}
The transformation between the two matrices corresponds to an exchange of the inter-block and intra-block positions, i.e., $\mathbf{H}_\mathrm{DD}^{i,j}[u,v] = \mathbf{H}^{u,v}[i,j]$. To illustrate this transformation, we highlight two elements \( a \) and \( b \) in both \( \mathbf{H}_\mathrm{DD} \) and \( \mathbf{H} \) in Fig.~\ref{two_H}. Element \(a\) is \(\mathbf{H}_\mathrm{DD}^{1,1}[5,2]\) in \(\mathbf{H}_\mathrm{DD}\) and maps to \(\mathbf{H}^{5,2}[1,1]\) after DDCP. The same mapping rule applies to element \(b\).

\begin{figure}
\centering
\includegraphics[width=3in]{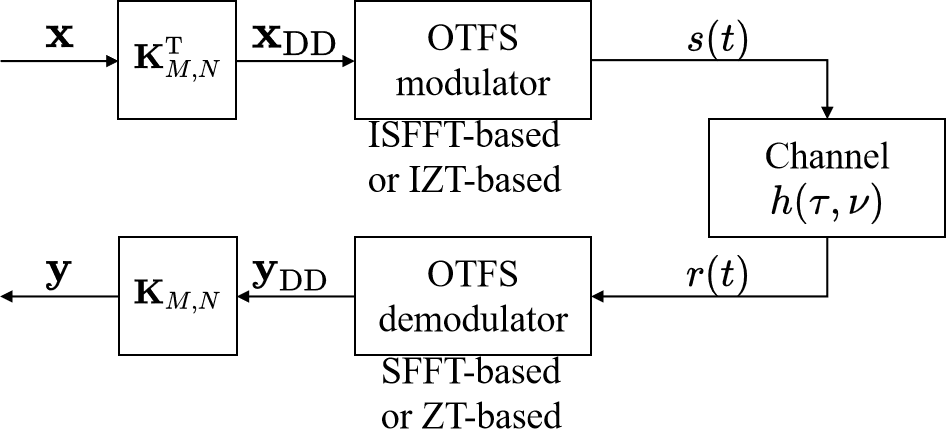}
\caption{The OTFS system model with DDCP for the transmitted and received symbol vectors.}
\label{OTFS_sys2}
\end{figure}

The OTFS system with DDCP is illustrated in Fig.~\ref{OTFS_sys2}. The precoder is given by $\mathbf{K}^{-1}_{M,N}=\mathbf{K}^\mathrm{T}_{M,N}$ before the OTFS modulator and the decoder is given by $\mathbf{K}_{M,N}$ after the OTFS demodulator. The overall input-output relation of this system with DDCP is given by (\ref{eq23}) instead of (\ref{eq4}). 

\section{Hybrid Message Passing Detection for OTFS Modulation}
In this section, we mainly consider the case of integer delay shifts, and our numerical results later will show that our detection scheme is also suitable for the case of fractional delay shifts. By applying the DDCP to the signal vector, we induce a structural transformation in the effective channel matrix, which exhibits inter-block sparsity. Based on this, we propose a hybrid MP algorithm for signal detection in OTFS systems. A linear estimator such as L-MMSE is applied to each dense block in channel $\mathbf{H}$. L-MMSE is well-suited for handling dense channel matrices because it performs linear estimation that accounts for interference from all symbols and does not rely on channel sparsity. In contrast, we employ the MP algorithm to suppress interference from other blocks in $\mathbf{H}$. The MP algorithm is well-suited for sparse channel matrices, as it operates efficiently on factor graphs with limited connectivity. By passing messages only along non-zero edges, MP significantly reduces computational complexity and avoids full-matrix operations. Traditional MP algorithms are typically designed for the case of integer delay and Doppler indices and will suffer from performance degradation in terms of BER and computational complexity with fractional delay and Doppler shifts. To overcome these limitations, the proposed hybrid scheme combines MP with an L-MMSE estimator, effectively exploiting the inter-block sparsity of the channel matrix \( \mathbf{H} \).\footnote{\label{foot:H}Recall that \( \mathbf{H} \) is the effective DD domain channel matrix by applying the DDCP to the signal vector  and it  can be partitioned into \( M \times M \) blocks of size \( N \times N \) as illustrated in Fig.~\ref{two_H}(b).}  
The proposed hybrid MP algorithm consists of the following two steps:
\begin{enumerate}
    \item First, we consider the segmentation of the symbol vector corresponding to the channel matrix partition. A hybrid detection approach is then employed: intra-block interference is suppressed using L-MMSE estimation, while inter-block interference is mitigated via the MP. 
    \item Second, a final symbol decision is performed based on the output of the hybrid MP detector using maximum \textit{a posteriori} (MAP) estimation.
\end{enumerate}
Among these two steps, the hybrid detection process in the first step is the core of the proposed algorithm. 

 \subsection{Step 1: Hybrid detection process}
To align with the block structure of \( \mathbf{H} \), we partition the transmitted and received symbol vectors \( \mathbf{x} \) and \( \mathbf{y} \) into \( M \) smaller sub-vectors, denoted by \( \mathbf{x}_m \in \mathbb{A}^{N} \) and \( \mathbf{y}_m \in \mathbb{A}^{N} \) for \( 1 \leq m \leq M \), respectively. Let \( \mathcal{I}(c) \) and \( \mathcal{J}(d) \) denote the sets of indices corresponding to the non-zero blocks in the \( c \)-th row and \( d \)-th column of \( \mathbf{H} \), respectively, with \( |\mathcal{I}(c)| = |\mathcal{J}(d)| = L \) for all rows and columns. \( L \) can also be regarded as the number of different delay shifts for paths in the wireless channel. Note that the parameter \(L\) is entirely determined by the channel characteristics, and no manual tuning or truncation is imposed. For channels with integer delay shifts, we have \(L \le P\) and \(L \le M\). When fractional delay indices are present, the effective delay shifts span the entire grid, and thus \(L = M\). Consequently, the input-output relation of the system can be expressed in a block-reduced form as
\begin{equation}
    \mathbf{y}_d = \sum_{f \in \mathcal{J}(d)} \mathbf{H}_{d,f} \mathbf{x}_f + \mathbf{n}_d,
    \label{eq24}
\end{equation}
where \( 1 \leq d \leq M \), \( 1 \leq f \leq M \), \( \mathbf{H}_{m_1, m_2} \) denotes the \( (m_1, m_2) \)-th block of the matrix \( \mathbf{H} \), and \( \mathbf{n}_d \) is an AWGN vector of length \( N \) with one-sided PSD \( N_0 \).

\begin{figure*}
\centering
\includegraphics[width=5in]{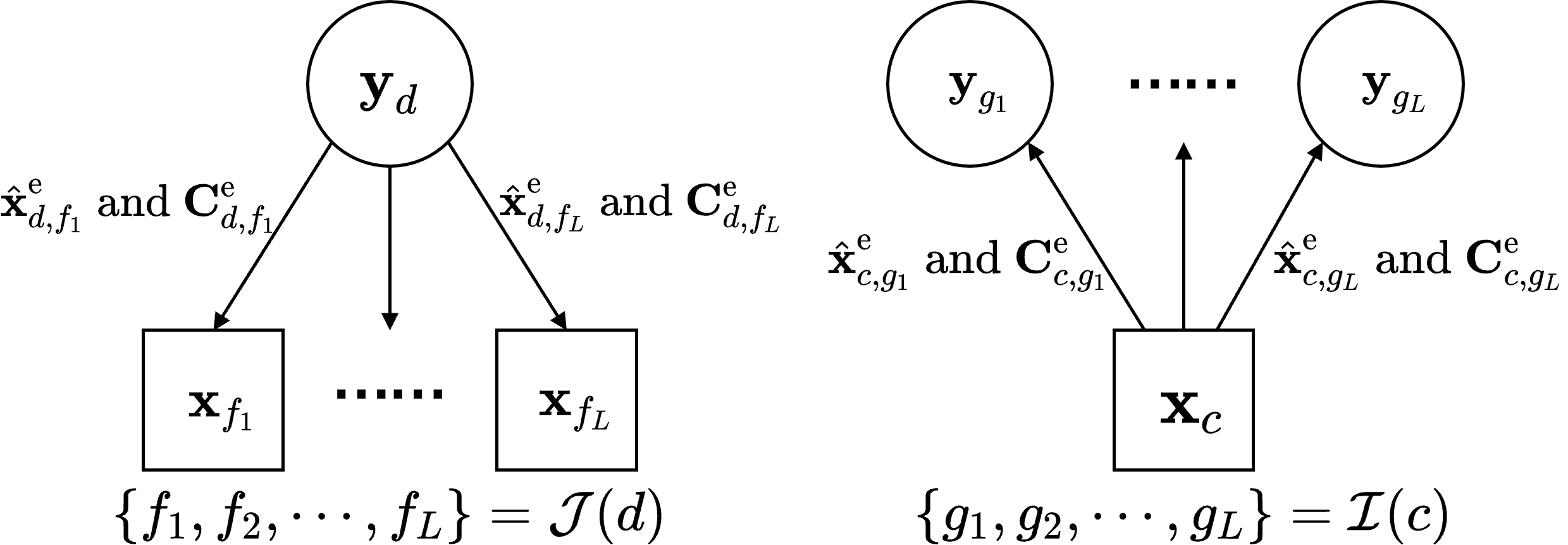}
\caption{Messages exchanged by OB $d$ and VB $c$ in the factor graph.  }
\label{factor_graph}
\end{figure*}

Based on (\ref{eq24}), the system can be modeled as a sparsely connected factor graph comprising \( M \) variable node blocks (VBs)  
and \( M \) observation node blocks (OBs). 
 The resulting factor graph is depicted in Fig.~\ref{factor_graph}. In this graph, each observation node block  \( \mathbf{y}_d \) (simply called OB $d$) is connected to a set of \( L \) variable node blocks \( \{ \mathbf{x}_f \mid f \in \mathcal{J}(d) \} \). Each variable node block \( \mathbf{x}_c \) (simply called VB $c$) is similarly connected to \( L \) observation node blocks \( \{ \mathbf{y}_g \mid g \in \mathcal{I}(c) \} \). It can be concluded that \( L \) corresponds to the number of distinct delay indices associated with the resolvable paths in the channel. 

 As illustrated in Fig.~\ref{factor_graph}, the detector comprises \(M\) VBs and \(M\) OBs, whose connections form \(ML\) VB--OB pairs.\footnote{\label{foot:neighboring}Note that each VB is connected to $L$ neighboring OBs.}  
All pairs follow the same update rule. 
The whole process contains multiple iterations, and 
in each iteration, each VB--OB pair proceeds synchronously with two sweeps. \textbf{Module A (Compute messages from OBs to VBs)}: each   OB  runs a local L-MMSE estimator in parallel to suppress intra-block interference using their local observations and the incoming information from the previous iteration; each OB then computes extrinsic messages and sends them once to its \(L\) neighboring VBs. \textbf{Module B (Compute messages from VBs to OBs)}: 
each VB 
processes in parallel by collecting the incoming messages from their connected OBs, combining them to refine symbol-wise beliefs. Meanwhile, a termination check is applied. If the termination criterion is not met, each VB computes and broadcasts updated extrinsic messages back to its $L$ connected OBs so as to reconcile cross-block interference, and the process moves to the next iteration; otherwise, the iterations stop, and final hard decisions are produced. 

\begin{figure*}[!t]
\centering
\includegraphics[width=7in]{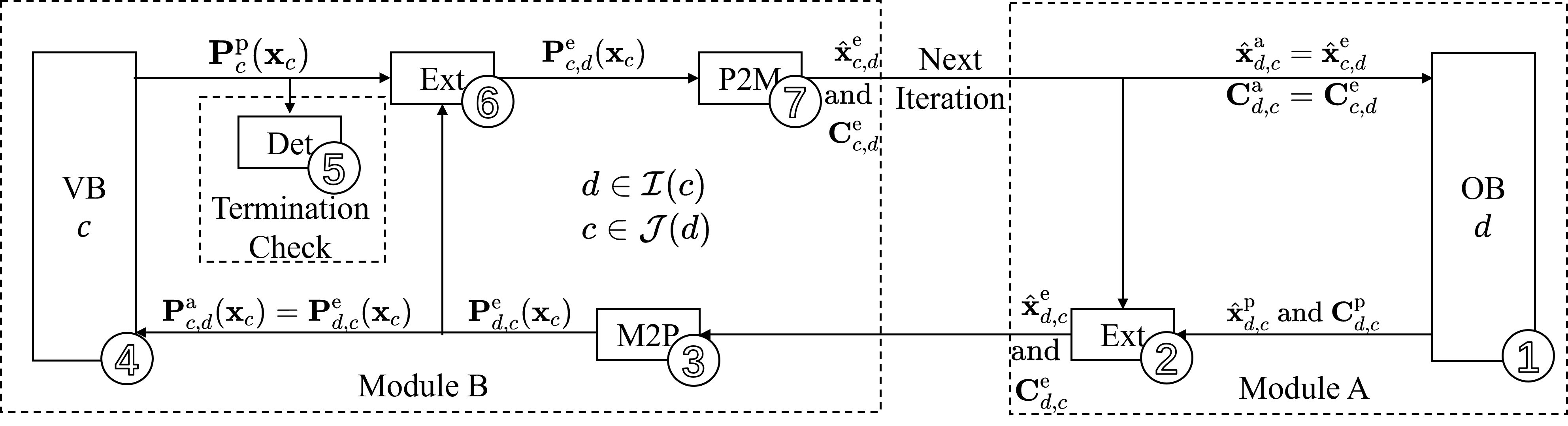}
\caption{The diagram corresponding to a single connection between VB $c$ and OB $d$ of the proposed OTFS detector.}
\label{diagram_det}
\end{figure*}

These two modules collectively comprise seven distinct operations, which are illustrated in Fig.~\ref{diagram_det}. For clarity, we focus on a single pair of VB \( c \) and OB \( d \) to elaborate on the details of the proposed iterative process as follows:
\begin{enumerate}
    \item \textbf{Module A:} {\it L-MMSE Estimation.} After obtaining OB \( d \), an L-MMSE estimator is applied to perform signal detection within a single block.
    \item \textbf{Module A:} {\it  Extrinsic Information Calculation (Matrix Form).} The ``Ext'' operation computes extrinsic information required for message passing, where the calculation is conducted in matrix form. 
    \item \textbf{Module B:} {\it  Matrix-to-Probability Transformation.} The ``M2P'' operation transforms the matrix-domain messages into element-wise probability distributions.
    \item \textbf{Module B:} {\it   Probability Combination.} This step aggregates all incoming messages at VB \( c \) to compute the \textit{a posteriori} element-wise probabilities of \( \mathbf{x}_c \).
    \item \textbf{Module B:} {\it   Detection.} The ``Det'' operation estimates \( \mathbf{x}_c \) based on the aggregated probabilities and checks for convergence using a predefined termination criterion. If the conditions are satisfied, the iterative process halts and the current estimate is output as the final detection result.
    \item \textbf{Module B:} {\it   Extrinsic Information Calculation (Probability Form).} A second ``Ext'' operation is performed, where extrinsic information is computed directly in the probability domain.
    \item \textbf{Module B:} {\it   Probability-to-Matrix Transformation.} The ``P2M'' operation converts the probability-domain messages back into matrix form for subsequent iterations.
\end{enumerate}

For notational simplicity, we introduce a set of concise symbols to represent the \textit{a priori}, \textit{a posteriori}, and extrinsic information associated with the estimated means, covariance matrices, and probability distributions of the partitioned OTFS signal vector \( \mathbf{x}_c \), as summarized in Table~\ref{tab:notations}. The subscript \( *_{c,d} \) denotes the message passed from VB \( c \) to OB \( d \), whereas the subscript \( *_{d,c} \) represents the message passed in the reverse direction, i.e., from OB \( d \) to VB \( c \).

\begin{table*}[!t]
\centering
\caption{Notations for Proposed Algorithm Parameters}
\label{tab:notations}
\begin{tabular}{|c|c|c|c|c|c|c|}
\hline
\multicolumn{1}{|c|}{} & \multicolumn{3}{c|}{VB $c$ to OB $d$} & \multicolumn{3}{c|}{OB $d$ to VB $c$} \\ \hline
& \textit{a priori} & \textit{a posteriori}  & extrinsic & \textit{a priori} & \textit{a posteriori}  & extrinsic  \\ \hline
estimates & N/A & N/A & $\mathbf{\hat{x}}^\mathrm e_{c,d}$, $\mathbf{C}^\mathrm e_{c,d}$ & $\mathbf{\hat{x}}^\mathrm a_{d,c}$, $\mathbf{C}^\mathrm a_{d,c}$ & $\mathbf{\hat{x}}^\mathrm p_{d,c}$, $\mathbf{C}^\mathrm p_{d,c}$ & $\mathbf{\hat{x}}^\mathrm e_{d,c}$, $\mathbf{C}^\mathrm e_{d,c}$  \\ \hline
probability  & $\mathbf{P}^\mathrm a_{c,d}(\mathbf{x}_c)$ & $\mathbf{P}^\mathrm p_c(\mathbf{x}_c)$ & $\mathbf{P}^\mathrm e_{c,d}(\mathbf{x}_c)$ & N/A & N/A & $\mathbf{P}^\mathrm e_{d,c}(\mathbf{x}_c)$  \\ \hline
\end{tabular}
\end{table*}

Before delving into the details, we highlight a key assumption: the symbols within each signal vector \( \mathbf{x}_c \) are assumed to be independent and identically distributed (i.i.d.) prior to the start of the iteration \cite{Ma2015}. Importantly, this statistical independence is maintained throughout the iterative process, even though each symbol may follow a distinct distribution as the iteration progresses. As a direct consequence, the covariance matrices illustrated in Fig.~\ref{diagram_det}—representing \textit{a priori}, \textit{a posteriori}, and extrinsic information—are constrained to be diagonal. To preserve this structure, the covariance matrices are computed on a symbol-wise basis, and all non-diagonal elements are explicitly discarded by setting them to zero. Therefore, only the diagonal entries are propagated and updated within the system.

Next we provide a more detailed description of Module A, Module B, and the termination criterion.


{\bf Module A: Compute Messages from OBs to VBs.}
For simplicity, our analysis primarily focuses on the iterative process between VB \( c \) and OB \( d \). The received signal at OB \( d \) is
\begin{equation}
\mathbf{y}_d =  \mathbf{H}_{d,c} \mathbf{x}_c+ \sum_{f \in \mathcal{J}(d), f \ne c} \mathbf{H}_{d,f} \mathbf{x}_f +   \mathbf{n}_d.
\label{eq25}
\end{equation}
The three components on the RHS of (\ref{eq25}) correspond to the desired signal from VB \( c \), interference from other VBs connected to OB \( d \), and AWGN, respectively. The objective of this step is to eliminate the interference term and then apply L-MMSE estimation to the desired signal. To this end, we utilize messages from VB \( f \), where \( f \in \mathcal{J}(d) \). The extrinsic mean and covariance from VB \( f \) are used as the \textit{a priori} statistics for \( \mathbf{x}_f \), i.e.,
\begin{equation}
\hat{\mathbf{x}}^\mathrm{a}_{d,f} = \hat{\mathbf{x}}^\mathrm{e}_{f,d}, \quad \mathbf{C}^\mathrm{a}_{d,f} = \mathbf{C}^\mathrm{e}_{f,d}.
\label{eq26}
\end{equation}
Using \( \hat{\mathbf{x}}^\mathrm{a}_{d,f} \), the interference can be eliminated, introducing an additional noise term:
\begin{equation}
\mathbf{y}_d - \sum_{f \ne c} \mathbf{H}_{d,f} \hat{\mathbf{x}}_{d,f}^\mathrm{a} = \mathbf{H}_{d,c} \mathbf{x}_c + \sum_{f \ne c} \mathbf{H}_{d,f} \left( \mathbf{x}_f - \hat{\mathbf{x}}_{d,f}^\mathrm{a} \right) + \mathbf{n}_d.
\label{eq27}
\end{equation}
We denote the aggregated noise term in (\ref{eq27}) as \( \mathbf{z}_{d,c} :=\mathbf{y}_d - \sum_{f \ne c} \mathbf{H}_{d,f} \hat{\mathbf{x}}_{d,f}^\mathrm{a} \), whose mean and covariance are given by
\begin{equation}
\mathbf{m}_{\mathbf{z}_{d,c}} = \mathbf{0}, \quad \mathbf{C}_{\mathbf{z}_{d,c}} = \sum_{f \ne c} \mathbf{H}_{d,f} \mathbf{C}_{d,f}^\mathrm{a} \mathbf{H}_{d,f}^\mathrm{H} + N_0 \mathbf{I}_N.
\label{eq28}
\end{equation}

{\bf Module A:} {\it L-MMSE Estimation.}
To estimate the symbol vector \( \mathbf{x}_c \), we employ the L-MMSE estimator in Module A, enhanced by the \textit{a priori} information from Module B. Note that \( \mathbf{C}^\mathrm{a}_{d,c} \) is assumed to be diagonal due to symbol independence and is initialized as \( \mathbf{I}_N \) in the first iteration. The corresponding L-MMSE estimation matrix is given by \cite{Sengijpta1993}
\begin{equation}
\mathbf{W}_{\mathrm{MMSE}} = \mathbf{C}_{d,c}^\mathrm{a} \mathbf{H}_{d,c}^\mathrm{H} \left( \mathbf{H}_{d,c} \mathbf{C}_{d,c}^\mathrm{a} \mathbf{H}_{d,c}^\mathrm{H} + \mathbf{C}_{\mathbf{z}_{d,c}} \right)^{-1}.
\label{eq29}
\end{equation}

The \textit{a posteriori} mean estimate of \( \mathbf{x}_c \) is then computed as
\begin{equation}
\hat{\mathbf{x}}_{d,c}^\mathrm{p} = \mathbf{W}_{\mathrm{MMSE}} \left( \mathbf{y}_d - \sum_{f \ne c} \mathbf{H}_{d,f} \hat{\mathbf{x}}_{d,f}^\mathrm{a} - \mathbf{H}_{d,c} \hat{\mathbf{x}}_{d,c}^\mathrm{a} \right) + \hat{\mathbf{x}}_{d,c}^\mathrm{a},
\label{eq30}
\end{equation}
and the corresponding \textit{a posteriori} covariance matrix is 
\begin{equation}
\mathbf{C}_{d,c}^\mathrm{p} = \mathbf{C}_{d,c}^\mathrm{a} - \mathbf{C}_{d,c}^\mathrm{a} \mathbf{H}_{d,c}^\mathrm{H} \left( \mathbf{H}_{d,c} \mathbf{C}_{d,c}^\mathrm{a} \mathbf{H}_{d,c}^\mathrm{H} + \mathbf{C}_{\mathbf{z}_{d,c}} \right)^{-1} \mathbf{H}_{d,c} \mathbf{C}_{d,c}^\mathrm{a}.
\label{eq31}
\end{equation}

{\bf Module A:}  {\it Extrinsic Information Calculation (Matrix Form).}
The final step in Module A involves the computation of extrinsic information. For effective iterative detection, it is essential to propagate extrinsic rather than the \textit{a posteriori} information between modules. Let \( \hat{\mathbf{x}}^\mathrm{e}_{d,c} \) and \( \mathbf{C}_{d,c}^\mathrm{e} \) denote the extrinsic mean and covariance, respectively. According to \cite{Sengijpta1993}, they are computed as
\begin{equation}
\mathbf{C}_{d,c}^\mathrm{e} = \left( \left( \mathbf{C}_{d,c}^\mathrm{p} \right)^{-1} - \left( \mathbf{C}_{d,c}^\mathrm{a} \right)^{-1} \right)^{-1},
\label{eq32}
\end{equation}
\begin{equation}
\hat{\mathbf{x}}_{d,c}^\mathrm{e} = \mathbf{C}_{d,c}^\mathrm{e} \left( \left( \mathbf{C}_{d,c}^\mathrm{p} \right)^{-1} \hat{\mathbf{x}}_{d,c}^\mathrm{p} - \left( \mathbf{C}_{d,c}^\mathrm{a} \right)^{-1} \hat{\mathbf{x}}_{d,c}^\mathrm{a} \right).
\label{eq33}
\end{equation}

Under the independence assumption, the updates in (\ref{eq32}) and (\ref{eq33}) can be further simplified to a symbol-wise form as
\begin{equation}
\mathbf{C}_{d,c}^\mathrm{e}[n,n] = \left( \frac{1}{\mathbf{C}_{d,c}^\mathrm{p}[n,n]} - \frac{1}{\mathbf{C}_{d,c}^\mathrm{a}[n,n]} \right)^{-1},
\label{eq34}
\end{equation}
\begin{equation}
\hat{\mathbf{x}}_{d,c}^\mathrm{e}[n] = \mathbf{C}_{d,c}^\mathrm{e}[n,n] \left( \frac{\hat{\mathbf{x}}_{d,c}^\mathrm{p}[n]}{\mathbf{C}_{d,c}^\mathrm{p}[n,n]} - \frac{\hat{\mathbf{x}}_{d,c}^\mathrm{a}[n]}{\mathbf{C}_{d,c}^\mathrm{a}[n,n]} \right),
\label{eq35}
\end{equation}
for \( 1 \le n \le N \). 

The pseudocode of the whole process of Module A is summarized in Algorithm~\ref{alg:moduleA}.

\begin{algorithm}
\caption{Calculation of the messages passed from OB $d$ to VB $c$}\label{alg:moduleA}
\renewcommand{\algorithmicrequire}{\textbf{Input:}}
\renewcommand{\algorithmicensure}{\textbf{Output:}}
\begin{algorithmic}[1]
\REQUIRE $\mathbf{y}_d$, $\mathbf{H}$, $\hat{\mathbf{x}}_{f,d}^\mathrm e$ and $\mathbf{C}_{f,d}^\mathrm e$ for $f \in \mathcal{J}(d)$.
\ENSURE $\hat{\mathbf{x}}_{d,c}^\mathrm e$ and $\mathbf{C}_{d,c}^\mathrm e$.
\FORALL{$f\in \mathcal{J}(d)$}
\STATE Get the \textit{a priori} mean and covariance matrix by (\ref{eq26}).
\ENDFOR 
\STATE Calculate the covariance matrix of noise $\mathbf{z}_{d,c}$ by (\ref{eq28}).
\STATE Compute the L-MMSE estimator matrix $\mathbf{W}_\text{MMSE}$ by (\ref{eq29}).

\STATE Calculate  the estimation output $\hat{\mathbf{x}}^\mathrm p_{d,c}$ by (\ref{eq30}) and the corresponding diagonal covariance matrix $\mathbf{C}^\mathrm p_{d,c}$ by (\ref{eq31}).

\STATE Compute the extrinsic output $\hat{\mathbf{x}}^\mathrm e_{d,c}$ and $\mathbf{C}^\mathrm e_{d,c}$ symbol-by-symbol based on (\ref{eq34}) and (\ref{eq35}).
\RETURN $\hat{\mathbf{x}}^\mathrm e_{d,c}$ and $\mathbf{C}^\mathrm e_{d,c}$.
\end{algorithmic}
\end{algorithm}

{\bf Module B: Compute Messages From VBs to OBs.}
In Module B, the first step is to perform the M2P operation, transforming matrix-form information into probability-form information. Specifically, we define a probability matrix \(\mathbf{P}_{d,c}^\mathrm e(\mathbf{x}_c)\) to characterize the likelihood that each symbol in \(\mathbf{x}_c\) takes a particular constellation point based solely on the observations at OB \( d \):
\begin{equation}
    \mathbf{P}_{d,c}^\mathrm e(\mathbf{x}_c)[n,q]=\mathrm{p}_{d,c}^\mathrm e(\mathbf{x}_c[n]=a_q).
    \label{eq36}
\end{equation}

Considering (\ref{eq25}), the joint maximum \textit{a posteriori} (MAP) detection rule for estimating the transmitted signals based on \(\mathbf{y}_d\) is given by
\begin{equation}
    \hat{\mathbf{x}}_{d,c} = \arg \max_{\mathbf{x}\in\mathbb{A}^N} \Pr \left( \mathbf{x} \mid \mathbf{y}_d,\mathbf{H} \right),
    \label{eq37}
\end{equation}
which has an exponential computational complexity in \(N\). As joint MAP detection is typically intractable for practical system dimensions \cite{Li20212}, and given the independence assumption among symbols within \(\mathbf{x}_c\), we adopt a symbol-wise MAP detection rule:
\begin{subequations}
\begin{align}
&\hat{\mathbf{x}}_{d,c}[n] \negmedspace = \negmedspace \arg \max_{a_q\in\mathbb{A}} \Pr\left( \mathbf{x}_c[n]=a_q  | \hat{\mathbf{x}}^\mathrm e_{d,c}[n], \mathbf{C}^\mathrm e_{d,c}[n,n] \right)
\label{eq38(1)} \\&=\arg \max_{a_q\in\mathbb{A}} \Pr\left( \hat{\mathbf{x}}^\mathrm e_{d,c}[n] \mid \mathbf{x}_c[n]=a_q, \mathbf{C}^\mathrm e_{d,c}[n,n] \right), 
\label{eq38}
\end{align}
\end{subequations}
where~\eqref{eq38(1)} originates from the L-MMSE estimator employed in Module A, and~\eqref{eq38} relies on the assumption of symbol independence. 
 The estimation error of \(\mathbf{x}_c[n]\) follows a normal distribution with zero mean and variance \(\mathbf{C}^\mathrm e_{d,c}[n,n]\). To simplify further expressions, we introduce an auxiliary function \(\xi(g,c,n,q)\) defined as follows for all \(g \in \mathcal{I}(c)\) \cite{Raviteja2018}:
\begin{equation}
    \xi(g,c,n,q)=\exp\left(\frac{-\left|\hat{\mathbf{x}}_{g,c}^\mathrm e[n]-a_q\right|^2}{\mathbf{C}_{g,c}^\mathrm e[n,n]}\right).
    \label{eq39}
\end{equation}

{\bf Module B:} {\it Matrix-to-Probability Transformation.}
Using \(\xi(g,c,n,q)\), the symbol-wise probability distribution of \(\mathbf{x}_c\) across constellation points can be expressed as
\begin{equation}
\begin{aligned}
\mathbf{P}_{d,c}^\mathrm e(\mathbf{x}_c)[n,q]  &\propto \Pr\left( \hat{\mathbf{x}}^\mathrm e_{d,c}[n] \mid \mathbf{x}_c[n]=a_q, \mathbf{C}^\mathrm e_{d,c}[n,n] \right)
\\&=\frac{\xi(d,c,n,q)}{\sum_{r=1}^Q\xi(d,c,n,r)}.
\label{eq40}
\end{aligned}
\end{equation}
However, directly computing (\ref{eq40}) is unnecessary. Instead, it suffices to calculate the values \(\xi(d,c,n,q)\) individually, preserving the proportional relationship:
\begin{equation}
\begin{aligned}
    \mathbf{P}_{d,c}^\mathrm e(\mathbf{x}_c)[n,1]:\cdots:\mathbf{P}_{d,c}^\mathrm e(\mathbf{x}_c)[n,Q] \\
    =\xi(d,c,n,1):\cdots:\xi(d,c,n,Q).
    \label{eq41}
    \end{aligned}
\end{equation}
The extrinsic probability matrix from OB \( d \) is treated as the \textit{a priori} probability matrix for VB \( c \), namely,
\begin{equation}
    \mathbf{P}_{c,d}^\mathrm a(\mathbf{x}_c) = \mathbf{P}_{d,c}^\mathrm e(\mathbf{x}_c).
    \label{eq42}
\end{equation}

{\bf Module B:} {\it Probability Combination.}
The next step involves combining the probabilities obtained from all OBs connected to VB \( c \) to derive the symbol-wise \textit{a posteriori} probability distribution of \( \mathbf{x}_c \) over the constellation points. In this step, the extrinsic information provided by OB \( g \), for each \( g \in \mathcal{I}(c) \), is treated as the \textit{a priori} information, denoted by \( \mathbf{P}^\mathrm{a}_{c,g}(\mathbf{x}_c) \). Specifically, \( \mathbf{P}^\mathrm{a}_{c,g}(\mathbf{x}_c) \) represents the likelihood that each symbol in \( \mathbf{x}_c \) takes a specific constellation point, based solely on the observation at OB \( g \).

Assuming mutual independence among the observations \( \mathbf{y}_g \) obtained from different OBs, the corresponding probabilities from these OBs can be combined multiplicatively. Therefore, the symbol-wise \textit{a posteriori} probability is computed as
\begin{equation}
\begin{aligned}
\mathbf{P}_{c}^{\mathrm p}(\mathbf{x}_{c})[n,q] &\propto\prod_{g\in \mathcal{I}(c)}\mathbf{P}_{c,g}^{\mathrm a}(\mathbf{x}_{c})[n,q] \\
&\propto \prod_{g\in \mathcal{I}(c)}\xi(g,c,n,q) \\
&=\frac{\prod_{g\in \mathcal{I}(c)}\xi(g,c,n,q)}{\sum_{r=1}^Q\prod_{g\in \mathcal{I}(c)}\xi(g,c,n,r)}.
\label{eq43}
\end{aligned}
\end{equation}
Before evaluating (\ref{eq43}), the function \(\xi(g,c,n,q)\) can be maintained in logarithmic form to improve numerical stability, reduce computational complexity, and mitigate the risk of numerical underflow during exponential calculations. At this stage, probability values must be explicitly computed to enable subsequent detection and to check the termination of the iteration process, where the termination criterion will be detailed later.

{\bf Module B:} {\it Extrinsic Information Calculation (Probability Form).}
Furthermore, if the iteration continues, the extrinsic probability \(\mathbf{P}_{c,d}^\mathrm{e}(\mathbf{x}_c)\) must be computed based on the \textit{a posteriori} probability \(\mathbf{P}_c^\mathrm{p}(\mathbf{x}_c)\). Since \(\mathbf{P}_c^\mathrm{p}(\mathbf{x}_c)\) is derived by combining observations from all OBs connected to VB \( c \), the extrinsic probability can be obtained by excluding the specific contribution from OB \( d \) \cite{Raviteja2018}. Specifically, we have
\begin{equation}
    \mathbf{P}_{c,d}^\mathrm{e}(\mathbf{x}_c)[n,q]=\frac{\prod_{g\neq d}\xi(g,c,n,q)}{\sum_{r=1}^Q\prod_{g\neq d}\xi(g,c,n,r)}, \forall{g\in\mathcal{I}(c)}.
    \label{eq44}
\end{equation}

To prevent abrupt changes in the extrinsic probabilities, which could potentially lead the iterative system into numerical instability or a singular state, we introduce a damping factor \(\Delta\). The updated extrinsic probability is thus computed as a weighted combination of the newly obtained probability and its previous iteration value:
\begin{equation}
[\mathbf{P}_{c,d}^\mathrm{e}(\mathbf{x}_c)]^{(i+1)} = \Delta \cdot  \mathbf{P}_{c,d}^\mathrm{e}(\mathbf{x}_c) + (1-\Delta)\cdot [\mathbf{P}_{c,d}^\mathrm{e}(\mathbf{x}_c)]^{(i)}.
\label{eq45}
\end{equation}
Note that (\ref{eq45}) implies that the extrinsic probability matrix at iteration \(i+1\) is obtained by mixing the newly computed result from (\ref{eq44}) and the extrinsic probability matrix from iteration \(i\), with weights \(\Delta\) and \(1-\Delta\), respectively. 

{\bf Module B:} {\it Probability-to-Matrix Transformation.}
The final procedure in Module B is the P2M operation, which converts the extrinsic probability information back into matrix form. Specifically, the extrinsic mean and covariance matrix of \(\mathbf{x}_c\) are computed based on its extrinsic probability distribution. By the assumption of symbol independence, these calculations can be carried out on a symbol-by-symbol basis as follows:
\begin{align}
  &  \hat{\mathbf{x}}_{c,d}^\mathrm e[n]=\sum_{q=1}^Q\mathbf{p}_{c,d}^\mathrm e(\mathbf{x}_c)[n,q] \cdot a_q,
    \label{eq46}\\
   & \mathbf{C}_{c,d}^\mathrm e[n,n]=\sum_{q=1}^Q\mathbf{p}_{c,d}^\mathrm e(\mathbf{x}_c)[n,q]\left|a_q-\hat{\mathbf{x}}_{c,d}^\mathrm e[n]\right|^2.
    \label{eq47}
\end{align}

The pseudocode of the whole process of Module B is summarized in Algorithm~\ref{alg:moduleB}. 
\begin{algorithm}
\caption{Calculation of the messages passed from VB $c$ to OB $d$}\label{alg:moduleB}
\renewcommand{\algorithmicrequire}{\textbf{Input:}}
\renewcommand{\algorithmicensure}{\textbf{Output:}}
\begin{algorithmic}[1]
\REQUIRE $\mathbb{A}$, $\hat{\mathbf{x}}^\mathrm e_{g,c}$ and $\mathbf{C}^\mathrm e_{g,c}$ for $g\in \mathcal{I}(c)$.
\ENSURE $\hat{\mathbf{x}}^\mathrm e_{c,d}$ and $\mathbf{C}^\mathrm e_{c,d}$ or $\mathbf{P}_m^\mathrm p(\mathbf{x}_m)$ for $1\le m\le M$.
\STATE Calculate $\xi(g,c,n,q)$ for $\forall{g\in \mathcal{I}(c)}$, $1\le n\le N$ and $1\le q\le Q$ by (\ref{eq39}).
\STATE Compute the \textit{a posteriori} probability matrix $\mathbf{P}^\mathrm p_c(\mathbf{x}_c)$ by (\ref{eq43}).
\STATE Check the termination criterion. 
\IF{The termination criterion is satisfied}
\RETURN $\mathbf{P}^\mathrm p_c(\mathbf{x}_c)$.
\ELSE
\STATE Renew the extrinsic probability $\mathbf{P}_{c,d}^\mathrm e(\mathbf{x}_c)$ by (\ref{eq44}) and (\ref{eq45}).
\STATE Calculate the extrinsic mean $\hat{\mathbf{x}}_{c,d}^\mathrm e$ and covariance $\mathbf{C}_{c,d}^\mathrm e$ based on (\ref{eq46}) and (\ref{eq47}). 
\RETURN $\hat{\mathbf{x}}_{c,d}^\mathrm e$ and $\mathbf{C}_{c,d}^\mathrm e$.
\ENDIF 
\end{algorithmic}
\end{algorithm}

{\bf Termination Criterion.}
Prior to calculating the extrinsic probability \(\mathbf{P}_{c,d}^\mathrm e(\mathbf{x}_c)\) in each iteration, we evaluate the \textit{a posteriori} probabilities obtained at VB \( c \) to determine whether the iterative loop should be terminated. To quantitatively assess the progress of signal detection across the entire OTFS symbol frame, we define a ratio \(\eta\) as follows:
\begin{equation}
\eta=\frac{1}{MN}\sum_{m=1}^M\sum_{n=1}^N\mathbb{I}\left(\max_q\mathbf{\mathrm P}_m^\mathrm p(\mathbf{x}_m)[n,q]\geq 1-\epsilon\right),
\label{eq48}
\end{equation}
where \(\epsilon<1\) is a predefined small positive threshold, and \(\mathbb{I}(*)\) denotes the indicator function which equals $1$ if $*$ is true, and equals $0$ otherwise.
The metric \(\eta\) represents the proportion of symbols within the OTFS frame that can be confidently estimated, i.e., the symbols whose probability of being correctly detected exceeds \(1-\epsilon\). Based on (\ref{eq48}), we define two rules for terminating the iteration process:
\begin{enumerate}
    \item {\it Early termination criterion:} If \(\eta\) reaches 1 during any iteration, we conclude that the iterative signal detection has converged. Consequently, the iterative loop terminates immediately, and the final detection decision is obtained using the probability matrices \(\mathbf{P}_m^\mathrm p(\mathbf{x}_m)\), \(1 \leq m \leq M\), from this last iteration.
    
    \item {\it Maximum iteration limit:} If the iteration count reaches a predefined maximum value \(I_\mathrm{max}\) without achieving \(\eta = 1\), the iterative loop terminates. In this case, the detection results are generated from the probability matrices \(\mathbf{P}_m^\mathrm p(\mathbf{x}_m)\), \(1 \leq m \leq M\),  from the final iteration.
\end{enumerate}

 \subsection{Step 2: Final symbol decision}
With the outputs of the hybrid MP algorithm, namely the probability matrices \(\mathbf{P}_m^\mathrm p(\mathbf{x}_m)\) for \(1 \le m \le M\), a final symbol decision is made based on the maximum probability criterion:
\begin{equation}
\mathbf{x}_m^\text{dec}[n] = {\arg \max_{a_q} \mathbf{P}_m^\mathrm p(\mathbf{x}_m)[n,q]}, \quad 1\le n \le N.
\label{eq50}
\end{equation}

\section{Performance analysis }
In this section, we first analyze the matched filter bound (MFB) of the OTFS communication system. The MFB serves as a theoretical lower bound on error performance, applicable to all conceivable detection algorithms; however, it is generally unattainable in practical systems \cite{Baas2001}. 
The computational complexity associated with each module of the proposed detection scheme is also studied. 

\subsection{MFB of the Considered OTFS System}
The MFB serves as a theoretical benchmark for evaluating the optimal BER performance of all detection algorithms. As discussed in Section II, we consider a delay-Doppler channel with \( P \) resolvable propagation paths, each associated with a complex channel coefficient \( h_p \). In this context, the MFB corresponds to the best-case scenario in which all inter-symbol interference is completely eliminated, and the receiver performs ideal coherent combining of all signal components. Under this interference-free condition, the system reduces to a set of independent parallel Gaussian channels, each symbol experiencing an effective channel gain determined by the coherent sum $\sum_{p=1}^P |h_p|^2$. This results in the ideal detection performance, reflecting the minimum achievable BER under perfect channel knowledge and optimal combining. The MFB is the numerical BER performance based on the input-output relation below: 
\begin{equation}
\mathbf{y}' = \left(\sum_{p=1}^P |h_p|^2\right) \mathbf{x} + \mathbf{z},
\label{eq51}
\end{equation}
where \(\mathbf{y}'\) denotes the received signal, \(\mathbf{x}\) is the transmitted signal vector, and \(\mathbf{z}\) represents the AWGN with one-sided PSD \( N_0 \). The MFB can be numerically evaluated through Monte Carlo simulations. Moreover, it can also be computed via well-known closed-form or quasi-closed-form methods \cite{MFB_compute,MFB_compute2}. Assuming that the path gains $\{h_p\}$ are i.i.d. random variables following $\mathcal{CN}(0,1/P)$ and QPSK modulation is employed (which are consistent with the simulation settings in the following section), we can derive the corresponding MFB in closed form as 
\begin{equation}
\begin{aligned}
\mathrm{MFB} =& 
\frac{1}{2\sqrt{\pi}}
\frac{\Gamma\!\left(P+\tfrac{1}{2}\right)}{\Gamma(P+1)}\\&
\left(\frac{2P}{\mathrm{SNR}}\right)^{P}
\, {}_2F_1\!\left(P,\,P+\tfrac{1}{2};\,P+1;\,-\tfrac{2P}{\mathrm{SNR}}\right),
\label{a1}
\end{aligned}
\end{equation}
where $\Gamma(*)$ denotes the Gamma function, and ${}_2F_1(a,b;c;z)$ is the Gauss hypergeometric function. The $\mathrm{SNR}$ in (\ref{a1}) is the signal-to-noise ratio of the QPSK system.

\subsection{Complexity Analysis }
The computational complexity of our proposed hybrid MP detection algorithm can be decomposed into two main parts, corresponding to the two distinct modules involved in the iterative process. We denote the number of iterations required to satisfy the termination criterion in (\ref{eq48}) as \(I_\text{iter}\). Consequently, Module A and Module B are each executed \(I_\text{iter} M L\) times during the entire detection procedure.

We first consider the computational complexity of a single execution of Module A at a given observation node. Initially, the interference from other variable nodes is eliminated using (\ref{eq27}), resulting in a complexity on the order of \(\mathcal{O}(L N^2)\). Subsequently, the covariance matrix of \(\mathbf{z}_{d,c}\) is computed via (\ref{eq28}), which incurs a complexity of \(\mathcal{O}(LN^3)\). The subsequent step involves performing the L-MMSE estimation based on the result from (\ref{eq27}), also leading to a computational complexity of \(\mathcal{O}(N^3)\). Finally, the extrinsic information is calculated symbol-wise from the \textit{a posteriori} information using (\ref{eq34}) and (\ref{eq35}), resulting in complexity \(\mathcal{O}(N)\).

Next, we analyze the computational complexity associated with a single execution of Module B. Initially, the values of \(\xi(g,c,n,q)\) are computed for \(g \in \mathcal{I}(c)\), \(1\le n \le N\), and \(1\le q \le Q\), as described by (\ref{eq39}), incurring a complexity of \(\mathcal O(L N Q)\). Calculating the \textit{a posteriori} probabilities using (\ref{eq43}) similarly results in complexity \(\mathcal O(L N Q)\). Subsequently, updating the extrinsic probabilities according to (\ref{eq44}) and (\ref{eq45}) yields complexity \(\mathcal O(L N Q)\). Finally, the computation of the extrinsic mean and covariance matrix using (\ref{eq46}) and (\ref{eq47}) again incurs complexity \(\mathcal O(N Q)\). 

Therefore, the total computational complexity per iteration of the proposed scheme is \(\mathcal{O}(L^2 M N^3 + L^2 M N Q)\), which scales polynomially with $M$ and $N$. In contrast, the near-optimal methods such as maximum likelihood (ML) or MAP detection typically exhibit exponential complexity, on the order of \(\mathcal{O}(Q^{M N})\), making them infeasible for large-scale OTFS systems (with large $M$ and $N$). Even the conventional L-MMSE detector requires inverting the entire channel matrix and thus incurs a cubic complexity of \(\mathcal{O}((M N)^3)\). Therefore, the proposed scheme achieves a favorable trade-off between computational efficiency and detection accuracy, offering a competitive and scalable solution for practical OTFS deployments under both sparse and dense channel conditions.

\section{ Numerical results }
The simulation settings are summarized as follows. The path gains \(\{h_p\}\) are i.i.d. \(\mathcal{CN}(0,1/P)\) so that the average channel power is normalized. Furthermore, delays and Dopplers are selected independently and uniformly. We set \(l_{\max}=8\) and \(k_{\max}=8\). For the integer delay case, the delay index \(l_p\) is drawn from \(\{0,1,\ldots,l_{\max}\}\). For the fractional delay case, \(l_p\) is drawn from the interval \([0,l_{\max}]\) (not restricted to integers). The Doppler index \(k_p\) is drawn uniformly from \([-k_{\max},k_{\max}]\). We use QPSK modulation with varying numbers of paths \(P\), and set \(M=32\) and \(N=16\). The maximum iteration count is \(I_{\max}=20\), the damping factor is \(\Delta=0.7\), and the convergence threshold is \(\epsilon=0.01\).

\subsection{Convergence Performance}
As our detection approach is iterative, evaluating its convergence behavior is essential. To this end, we employ two complementary metrics: (i) the cumulative distribution function (CDF) of the mean squared error (MSE) associated with the extrinsic messages exchanged between the variable nodes and the observation nodes,\footnote{For brevity, we report results for integer-delay channels, and the fractional-delay case exhibits the same qualitative behavior.}
  and (ii) the number of iterations required to satisfy the predefined termination criterion. 

\begin{figure*}
\centering
\subfloat[CDFs with $P=2$]{\includegraphics[width=3in]{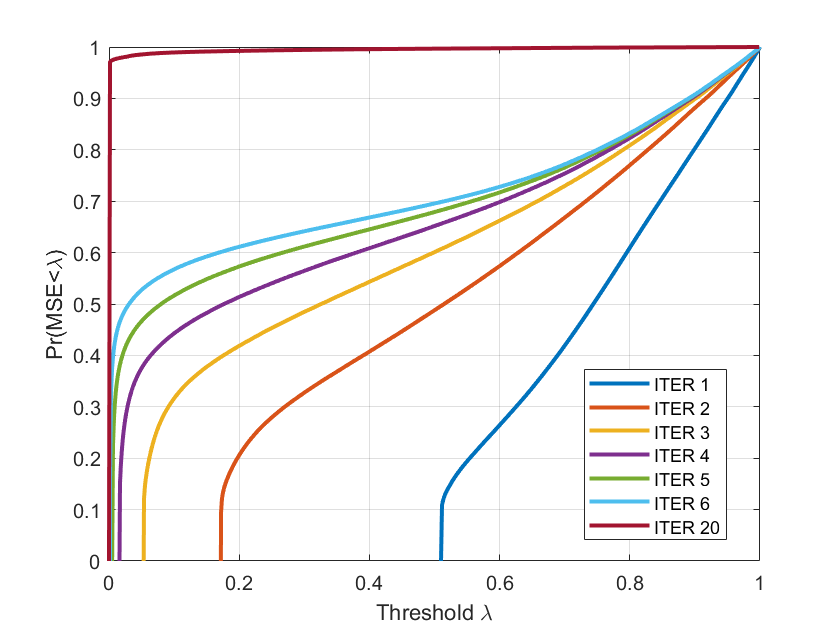}}%
\hfil
\subfloat[CDFs with $P=3$]{\includegraphics[width=3in]{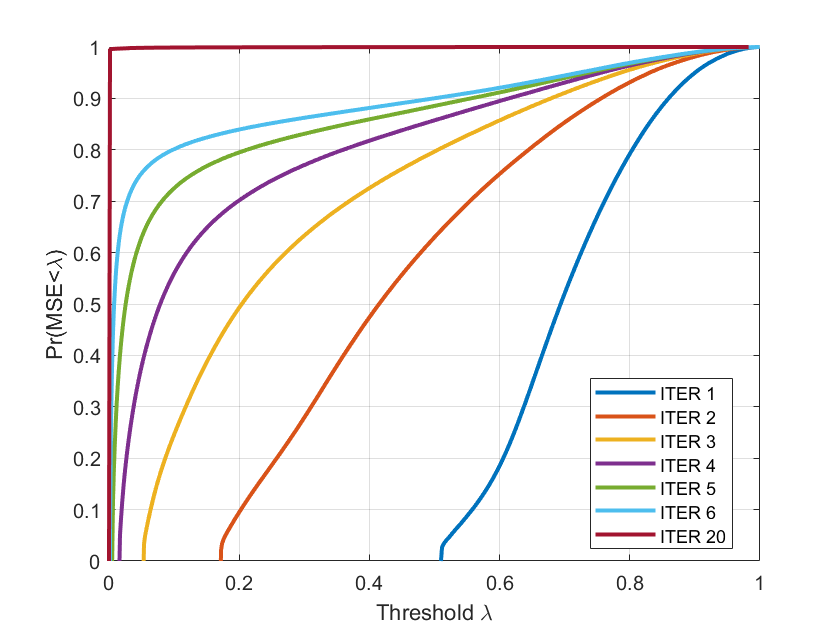}}%
\hfil
\subfloat[CDFs with $P=4$]{\includegraphics[width=3in]{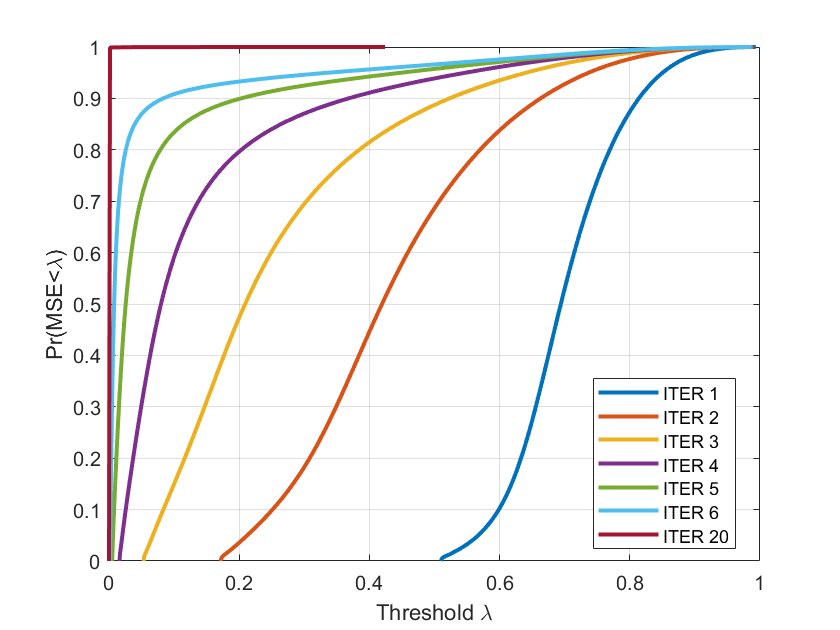}}%
\hfil
\subfloat[CDFs with $P=6$]{\includegraphics[width=3in]{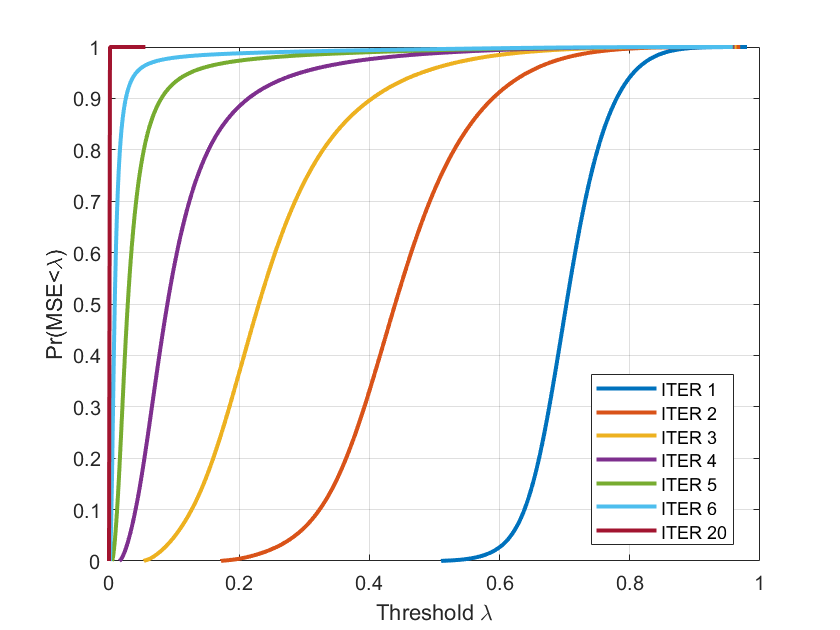}}%
\caption{CDFs of average MSE with different $P$ and ITER.} 
\label{cdf}
\end{figure*}

We first evaluate the CDF of the average MSE for the extrinsic messages sent from variable node \(c\) to observation node \(d\), denoted by \(\hat{\mathbf{x}}^{\mathrm e}_{c,d}\), for \(1\le c\le M\) and \(d\in\mathcal{I}(c)\). To reveal the intrinsic behavior, early termination is disabled in this study. The CDF is easy to read: a curve closer to the upper-left corner indicates a higher proportion of messages with low MSE, corresponding more closely to the true symbols. Fig.~\ref{cdf} shows the simulation results on this evaluation. For each subfigure (with fixed \(P\)), the curves move toward the upper-left as the iteration count increases and then settle, which indicates convergence as the message MSE decreases with more iterations. At a fixed iteration budget (e.g., \(\mathrm{ITER}=6\)), increasing \(P\) shifts the CDFs upward (see \(P=2,3,4,6\)). With more paths, the extrinsic information clusters more tightly around the constellation points, so the decisions become more accurate. This advantage of larger \(P\) will also be reflected in the BER results.

\begin{figure}
\centering
\includegraphics[width=3.5in]{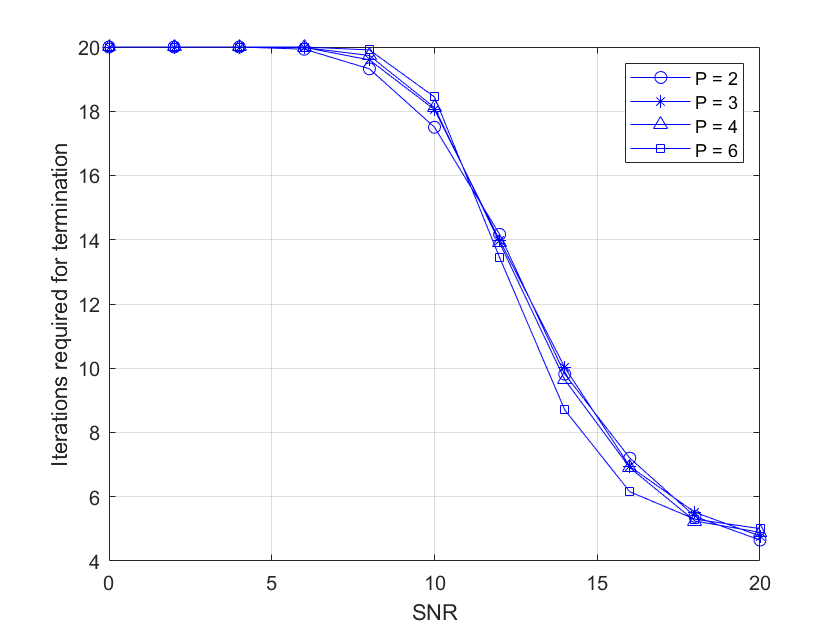}
\caption{The number of iterations required to meet the termination criterion with different $P$ for the integer delay case.}
\label{iter_int}
\end{figure}

\begin{figure}
\centering
\includegraphics[width=3.5in]{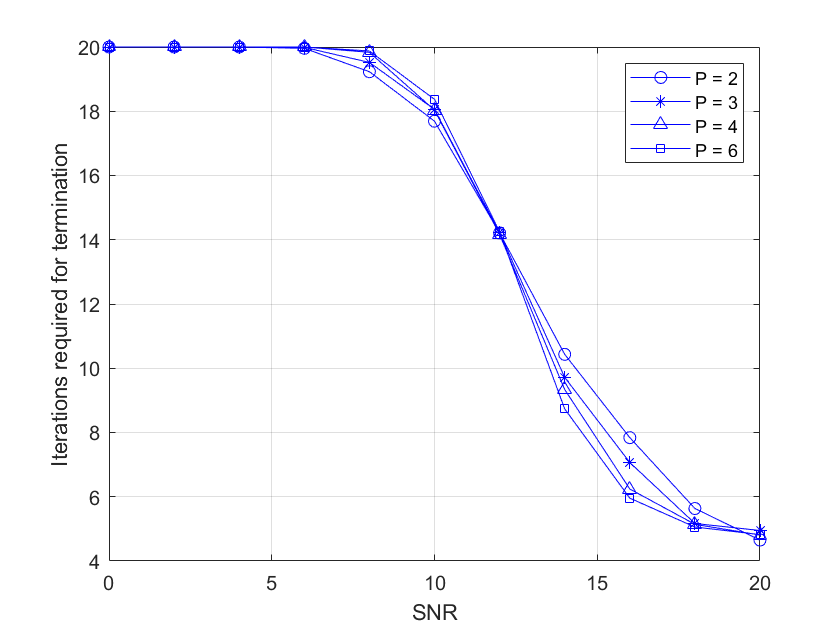}
\caption{The number of iterations required to meet the termination criterion with different $P$ for the fractional delay case.}
\label{iter_fra}
\end{figure}

Furthermore, we investigate the average number of iterations required for convergence under varying path numbers \(P\). The simulation results with integer delay shifts are given in Fig.~\ref{iter_int} and the results with fractional ones are shown in Fig.~\ref{iter_fra}. The two figures show the same qualitative behavior. The curves for different \(P\) are nearly overlapping, indicating that the required iterations are largely insensitive to \(P\). At low SNR (below about \(6\,\mathrm{dB}\)), the average count approaches the cap \(I_{\max}\), meaning that few frames satisfy the convergence criterion in (\ref{eq43}) within the allowed limit. As SNR increases from roughly \(8\) to \(18\,\mathrm{dB}\), the average count drops sharply from about \(20\) to around \(5\), showing much faster convergence. Beyond \(18\,\mathrm{dB}\), it stabilizes near 5 iterations.

In conclusion, as \(P\) increases, the MSE of extrinsic messages converges more tightly, while the number of iterations required to satisfy the early-termination criterion remains insensitive.

\subsection{BER performance}
In this part, we evaluate the BER performance of our proposed hybrid MP detection scheme under scenarios involving both integer and fractional delay indices. For comparison, we also present the BER performance of the L-MMSE detector, based on the DD domain effective channel matrix \(\mathbf{H}_\mathrm{DD}\), along with the MFB, which serves as a theoretical benchmark representing approximate optimal performance.

\begin{figure}
\centering
\includegraphics[width=3.5in]{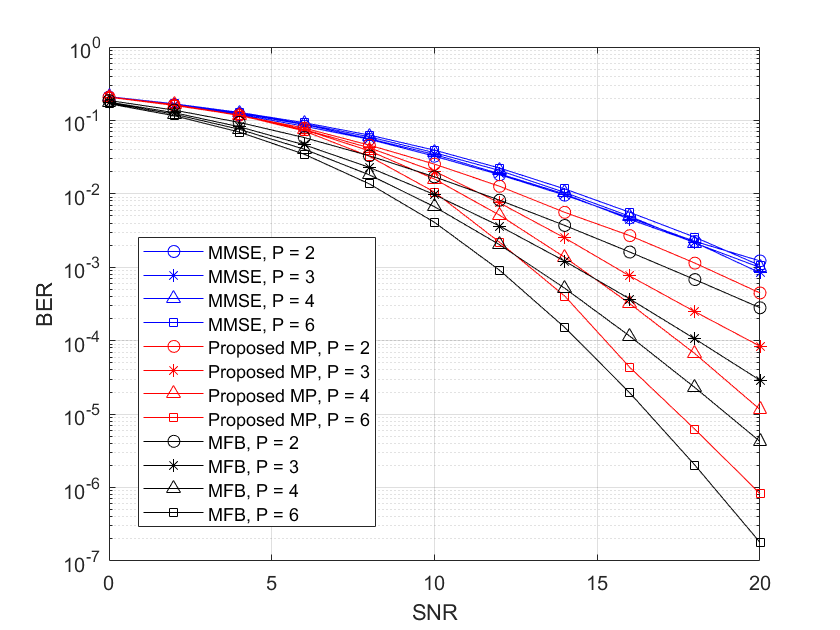}
\caption{BER performance for OTFS modulation with integer delay shifts, including L-MMSE, MP, and MFB.}
\label{ber_int}
\end{figure}

We first consider integer delay channels under the ISFFT/SFFT model: for each resolvable path, the delay index is drawn as an integer in \([0, l_{\max}]\) according to (\ref{eq12}) and (\ref{eq20}). Fig.~\ref{ber_int} compares the BER of the full-size L-MMSE detector, the proposed hybrid MP detector, and the MFB for different numbers of paths \(P\). The L-MMSE curves are nearly insensitive to \(P\), indicating limited exploitation of multipath diversity and degrees of freedom (DoF). In contrast, the proposed hybrid MP detector closely approaches the MFB in multipath scenarios, and its high-SNR slope tracks the bound, indicating near-optimal error performance and no noticeable DoF loss. At a BER of about \(10^{-3}\), the SNR gains over L-MMSE are approximately \(2.1\), \(4.1\), \(5.6\), and \(7.2\) dB for \(P=2\), \(3\), \(4\), and \(6\), respectively. Hence, the advantage of the proposed scheme grows with \(P\), consistent with our earlier CDF results: as the number of paths increases, the MSE of extrinsic messages concentrates closer to zero, improving detection accuracy. The computational cost also increases with \(P\), since a larger number of paths typically leads to a larger \(L\).

\begin{figure}
\centering
\includegraphics[width=3.5in]{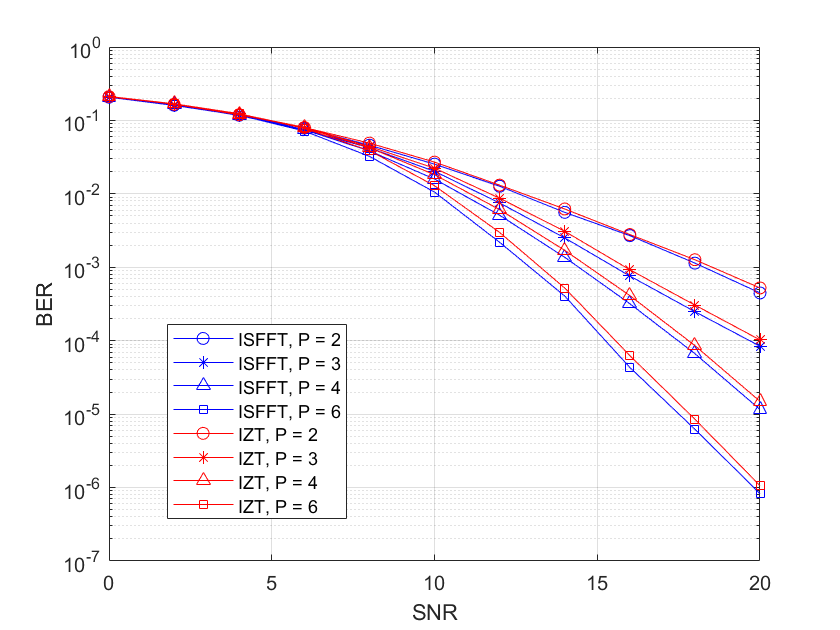}
\caption{BER performance of hybrid MP algorithm with different models for channel generation.}
\label{ber_isfftizt}
\end{figure}

Next, we compare the BER performance of our proposed detection scheme under two different OTFS channel models: the ISFFT/SFFT-based model and the IZT/ZT-based model, considering integer delay shifts and fractional Doppler shifts, as illustrated in Fig.~\ref{ber_isfftizt}. It can be observed that although the BER performance with the IZT/ZT model is slightly inferior to that of the ISFFT/SFFT model, the overall trends of both performance curves remain highly consistent. This demonstrates that our proposed hybrid MP algorithm is effective and achieves comparable BER performance for OTFS systems under both implementation schemes.

\begin{figure}
\centering
\includegraphics[width=3.5in]{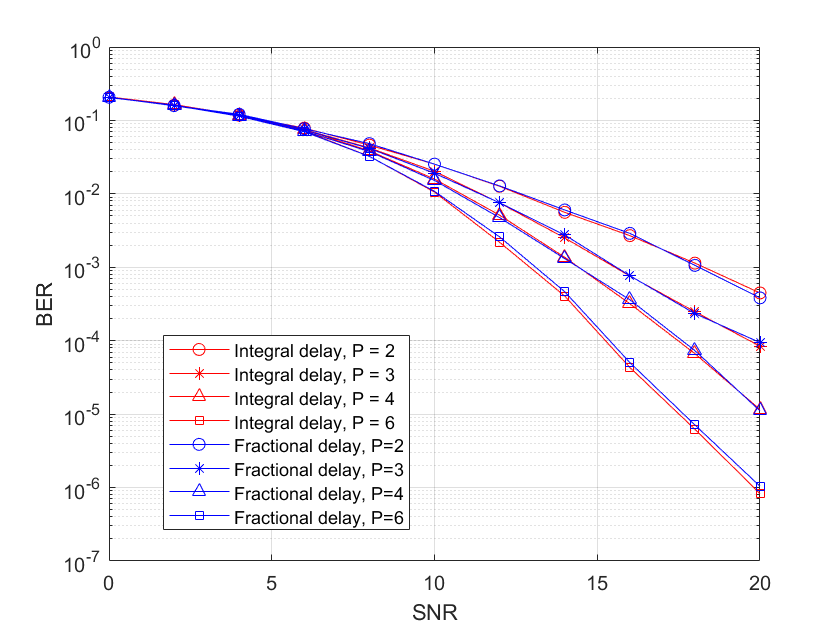}
\caption{BER performance of proposed hybrid MP algorithm, in the case of both integer and fractional delay indices.}
\label{ber_fra}
\end{figure}

Finally, we investigate the fractional delay index scenario. The effective channel matrix under fractional delay indices is generated using the IZT and ZT. The fractional delay indices for each resolvable path are randomly chosen from the continuous interval \([0, l_\mathrm{max}]\). Fig.~\ref{ber_fra} compares the BER performance under both integer and fractional delay scenarios for various numbers of resolvable paths \(P\). Remarkably, it is evident from the figure that the BER performance under fractional delay indices precisely matches that observed under integer delay indices. In other words, although fractional delay indices increase the computational complexity of the proposed detection algorithm as \(L=M\), they do not negatively affect the achievable BER when the proposed algorithm is applied.

In conclusion, the proposed hybrid MP detector outperforms the full-size L-MMSE method and closely approaches the MFB in multipath scenarios, with no discernible DoF loss. Moreover, it achieves similar BER under both integer and fractional delays and under both ISFFT/SFFT- and IZT/ZT-based channel models.

\section{Conclusion}
This paper proposed a hybrid iterative detection algorithm for OTFS systems by combining L-MMSE estimation with MP-based probabilistic inference. By applying the DDCP, the scheme effectively separates and mitigates intra- and inter-block interference. It demonstrated strong robustness under both sparse and dense DD channels and could be applied to both ISFFT/SFFT and IZT/ZT-based implementations. Simulation results confirmed that the proposed method significantly improves BER performance and approaches the MFB while maintaining manageable complexity. As ongoing work, we further extend this framework and develop complexity-reduction techniques for scenarios involving both fractional delay and Doppler shifts.

\bibliographystyle{IEEEtran}
\bibliography{note}

\end{document}